\documentclass[12pt,a4paper]{cas-sc}

\ExplSyntaxOn
\cs_gset:Npn \__first_footerline:
  { \group_begin: \small \sffamily \__short_authors: \group_end: }
\ExplSyntaxOff

\usepackage[numbers]{natbib}
\usepackage{setspace}
\usepackage{chemformula} 
\usepackage[T1]{fontenc} 
\usepackage{caption}
\usepackage{float}
\usepackage{geometry}
\usepackage{graphicx}
\usepackage{xkeyval}
\usepackage{svg}
\usepackage{subcaption}
\usepackage{wrapfig}
\usepackage{float}
\usepackage{amsmath}
\usepackage{bm}
\usepackage{array}
\usepackage[ruled,linesnumbered]{algorithm2e}

\def\tsc#1{\csdef{#1}{\textsc{\lowercase{#1}}\xspace}}
\tsc{WGM}
\tsc{QE}

\begin{document}
	
\title[mode=title]{Conditional grain-graph diffusion for property-guided inverse design of polycrystalline microstructures}

\author{Yuheng Zhou}[type=editor,
auid=000,bioid=1,orcid=0000-0002-2001-718X]
\credit{Conceptualization, Methodology, Formal analysis, Investigation, Data curation, Validation, Visualization, Writing - original draft}

\author{Xiao Shang}[type=editor,
auid=000,bioid=1,orcid=0009-0007-0888-5746]
\credit{Conceptualization, Investigation, Data curation, Writing - review and editing}

\author{Huicong Chen}[type=editor,auid=001,
orcid=0000-0001-8240-3193]
\credit{Investigation, Writing - review and editing}

\author{Yu Zou}[type=editor,auid=001,
orcid=0000-0002-0179-3642]
\ead{mse.zou@utoronto.ca}
\cormark[1]
\credit{Conceptualization, Supervision, Funding acquisition, Writing - review and editing}

\affiliation{organization={Department of Materials Science and Engineering, University of Toronto},
    city={Toronto ON},
    postcode={M5S 3E4},
    country={Canada}}

\shortauthors{Zhou et al.}
\shorttitle{}

\cortext[cor1]{Corresponding author}
\let\WriteBookmarks\relax
\def\floatpagepagefraction{1}
\def\textpagefraction{.0001}

\begin{abstract}[SUMMARY]
    \setstretch{1.2}
    
    Graph representations compactly encode polycrystalline microstructures while retaining grain topology and grain boundary information. This work presents a conditional graph diffusion framework for the property-guided inverse design of dual-phase Ti-6Al-4V microstructures. An enhanced grain graph neural network (GNN), incorporating grain boundary edge features, learnable node and edge embeddings, and multi-statistic pooling, serves as a forward surrogate for stress prediction and candidate evaluation. A conditional graph diffusion model subsequently generates grain microstructure candidates via reverse diffusion under prescribed \(\alpha\)-phase volume fraction, elastic modulus, and yield-stress proxy targets. Evaluations across four target regimes and independently seeded starting sets show that the generated candidates consistently approach the prescribed properties, including a target beyond the property envelope of the existing microstructures. For Ti-6Al-4V specifically, local crystallographic consistency is quantified post-generation based on deviations from the Burgers orientation relationship (BOR). BOR-aware ranking partially recovers the crystallographic consistency, producing relative increases in mean BOR consistency of up to 44.9\% and 56.4\% for the in- and out-of-envelope targets, respectively, while retaining close property-target alignment. Selected candidates are further validated by finite element analysis, with the mean properties of the five best candidates in each primary design case exhibiting a maximum absolute relative error of 1.0\%. In a representative benchmark, the diffusion model uses 32 candidate evaluations per input graph versus approximately 40,000 for random search and evolutionary optimization, while reducing runtime by approximately two orders of magnitude in the tested implementations. These results establish conditional grain-graph diffusion as an efficient framework for property-guided polycrystalline microstructure design.
    
    \end{abstract} 

\maketitle
\doublespacing

\section{Introduction}
The macroscopic mechanical response of polycrystalline metallic materials is strongly controlled by grain-level microstructural features, including grain size, grain morphology, phase distribution, and crystallographic orientation. These features govern local deformation responses and collectively influence effective properties such as elastic modulus, yield strength, ductility, fatigue resistance, and creep performance \cite{FIGUEIREDO20243448,zhu2026mechanical,Xu2026HighImpactResistance}. Additive manufacturing (AM) can produce heterogeneous grain microstructures that differ substantially from those obtained by conventional processing, while also providing increasing control over grain morphology, texture, and phase distribution \cite{gao2023additive,SHANG2026121952}. As a result, the process-structure-property relationship in AM metals provides both a challenge and an opportunity for microstructure level materials design \cite{WEI2025133,zhu2026mechanical}. Recent advances in metal AM have provided increasingly effective routes for controlling grain morphology, texture, and phase distribution. For example, grain structures can be manipulated through laser beam shaping, ultrasound-assisted solidification, and alternating process parameters during directed energy deposition \cite{ROEHLING2020109071, TODARO2021101632,ZHANG2022102865,SHANG2026105240}. These process control strategies expand the attainable microstructure space and provide practical pathways for tailoring grain structures, making grain-level inverse design increasingly relevant for application specific materials development.

High-fidelity numerical simulations provide a direct route for evaluating the mechanical response of candidate polycrystalline microstructures. Typically, representative volume elements are generated from grain microstructures, meshed, and evaluated under prescribed loading conditions using finite element analysis (FEA) \cite{KNEZEVIC2014239,QUEY20111729}. However, because polycrystalline microstructures contain stochastic grain morphologies, orientations, and phase distributions, iterative microstructure design requires repeated evaluation of many candidate structures. Direct high-fidelity simulation therefore becomes computationally expensive when used as the property evaluator inside an inverse design loop. Machine learning-based surrogate models provide an efficient alternative by learning the mapping from microstructure to properties using a limited set of high-fidelity simulation labels. Once trained, these models can rapidly evaluate unseen microstructures and can therefore be embedded into inverse design workflows. However, the accuracy, scalability, and physical interpretability of these surrogates depend strongly on the microstructure representation \cite{BOSTANABAD20181}.

Existing microstructure representations can be broadly grouped into statistical descriptor-based, image- or voxel-based, and graph-based approaches. Statistical descriptor-based methods represent microstructures using compact features such as correlation functions, texture descriptors, or reduced-order embeddings \cite{BOSTANABAD20181,HASHEMI2021110132}. These approaches are computationally efficient, but they can compress away grain microstructure information. Image- and voxel-based representations retain spatially resolved morphological information and are naturally compatible with convolutional neural networks (CNN) \cite{SHANG202371,IBRAGIMOVA2022103374,LIU2022108332}. Nevertheless, voxel-based representations scale poorly with 3D resolution and domain size, and grain boundaries are not represented as explicit physical entities. Graph-based representations provide a natural alternative for polycrystalline materials. In a grain graph, grains are represented as nodes and grain boundaries as edges, allowing grain-level features such as phase, volume, and crystallographic orientation to be combined with boundary-level features such as connectivity and boundary area. When paired with graph neural networks (GNN), these representations have been shown to provide compact and physically interpretable descriptions of polycrystalline microstructures \cite{Dai2021,DAI2023112461}. Graph-based models have been used for effective property prediction in polycrystalline materials and grain evolution or tracking \cite{QIN2024113061,GAO2025117844,VENKATESH2026121885}. In closely related work on dual-phase Ti-6Al-4V microstructures, Gao et al. combined a graph attention network (GAT) for predicting global stress-strain responses with a 3D conditional denoising diffusion probabilistic model for predicting local stress field distributions \cite{GAO2025117844}. The framework therefore provides forward predictions at both global and local scales. Importantly, however, its diffusion model predicts stress fields for prescribed microstructures rather than generating new grain microstructures under specified property targets.

The availability of fast surrogate models has accelerated inverse materials design. Search based and evolutionary optimization methods have been widely used to explore material and microstructure design spaces \cite{SHANG202371,D4MH00337C,D1MH01792F,Kim2021}. These approaches can be extended to multi-objective design through Pareto-based selection or scalarized objective functions. However, they still require many repeated candidate evaluations and become increasingly challenging as the number of design variables grows \cite{LIU2024101466}. This limitation is especially important for grain-level design, where each microstructure may contain many phase and orientation variables. Generative models provide a complementary route by learning the distribution of feasible designs and generating candidates directly near a desired response window. Flow-based generative models have recently been explored for property-guided microstructure inverse design \cite{MIRZAEE2025120704}. Among generative approaches, diffusion models are particularly well suited for inverse microstructure design because they learn a stochastic reverse denoising process and can generate diverse candidates conditioned on target properties. Recent studies have applied diffusion models to the inverse design of microstructure morphologies and topologies \cite{VLASSIS2023116126,Xu2026StyleConstrained} and various classes of metamaterials and architected structures \cite{Bastek2023,WANG2024117440,Yang2025GraphDGM,Zhang2025LatticeOptDiff,Zheng2026}. For polycrystalline materials, diffusion models have also been explored for statistically conditioned microstructure generation and for inverse prediction of process parameters and dendritic microstructures from mechanical properties \cite{BUZZY2024119746,FERNANDEZZELAIA2024101976,Ikeda2025}.

Overall, prior studies have established grain-graph models for efficient forward grain microstructure property prediction, whereas diffusion-based inverse design has been demonstrated primarily for image- and voxel-based microstructures. The direct application of conditional diffusion to polycrystalline grain graphs for generative microstructure design therefore remains largely unexplored. To address this gap, this work develops a conditional grain-graph denoising diffusion probabilistic model (DDPM) for property-guided inverse design of dual-phase Ti-6Al-4V microstructures. The diffusion model is trained to generate grain microstructure designs under prescribed elastic modulus, yield-stress proxy, and \(\alpha\)-phase volume fraction targets. An enhanced grain GNN forward surrogate incorporating grain boundary features, learnable feature embeddings, and multi-statistic graph pooling provides property prediction and candidate evaluation.  Selected candidates are subsequently reconstructed and validated by FEA. Two primary target regimes are considered: an in-envelope design-to-spec case motivated by biomedical applications requiring reduced stiffness while retaining sufficient strength \cite{Safavi2023ControlledPorousOrthopedic}, and an out-of-envelope extreme-property case, defined relative to the existing microstructures, motivated by aerospace applications requiring high stiffness and strength \cite{Gradl2022RobustMetalAM}. Sensitivity to starting-set selection and target location is further examined across four target regimes using three independently seeded sets of initial microstructures. Local Burgers orientation relationship (BOR) consistency is quantified post-generation, and BOR-aware ranking is introduced to balance property-target alignment with crystallographic consistency. Finally, the target-matching performance and computational efficiency of the trained diffusion framework are further compared with random search and evolutionary optimization. 

\section{Polycrystalline Microstructure Database}

\begin{figure}[ht]
\begin{center}
 \centering
 \includegraphics[width=400pt]{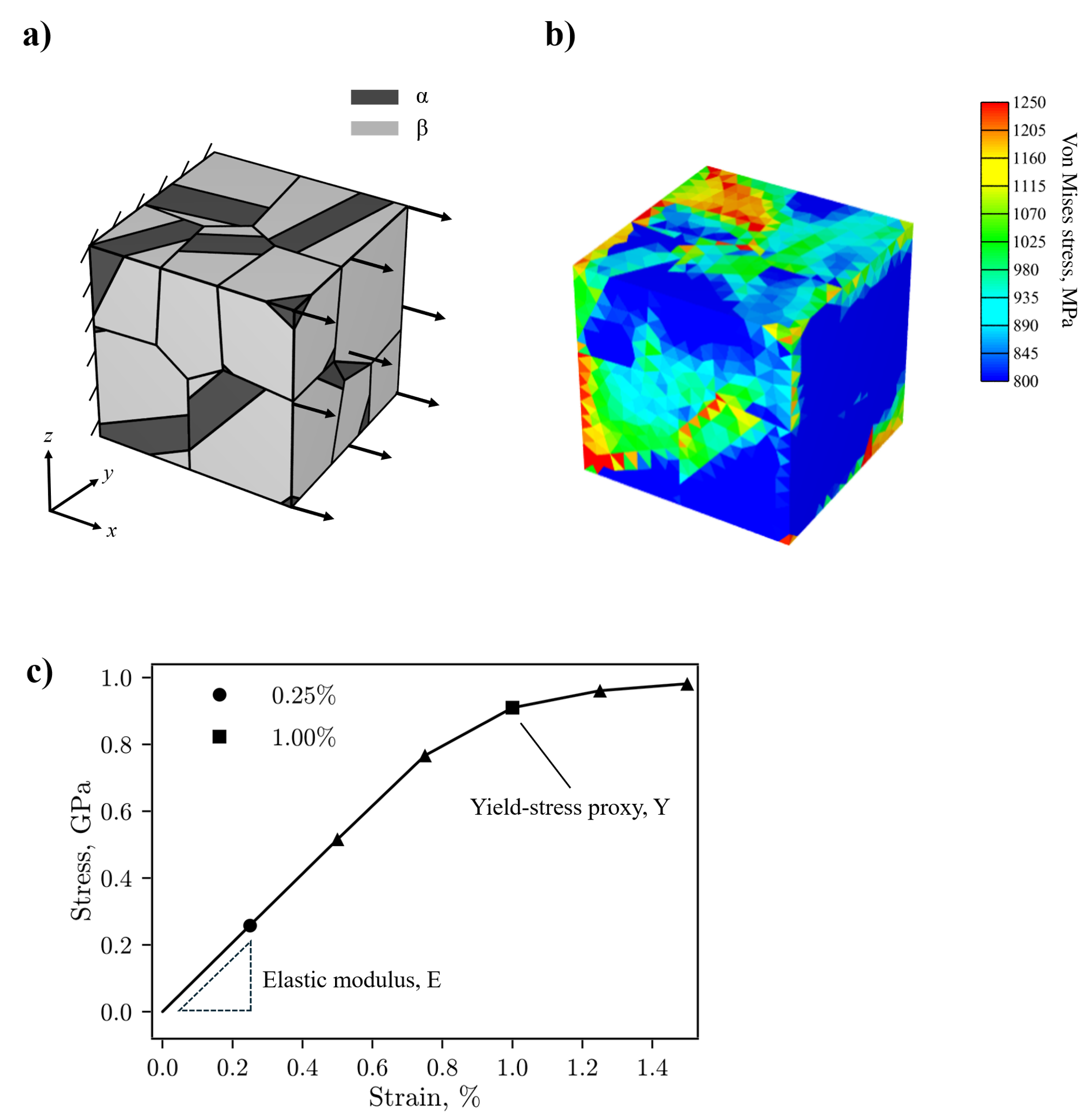}
\end{center}
\vspace{-1em}
\caption{Finite element simulation setup and property extraction from the Ti-6Al-4V microstructure database: a) Schematic of the uniaxial tensile loading boundary conditions. b) Representative von Mises stress distribution at the final imposed strain of 1.5\% c) Representative simulated stress-strain response, elastic modulus and yield-stress proxy extracted from the stress values at 0.25\% and 1.00\% strain, respectively.}
\label{fig:FEASetup}
\end{figure}

For the development and demonstration of the proposed polycrystalline microstructure design pipeline, 5,177 dual-phase Ti-6Al-4V grain microstructures are derived from the microstructure database developed by Shang et al. \citep{SHANG202371}. As illustrated in Fig. \ref{fig:FEASetup} a), the \(\alpha+\beta\) Ti-6Al-4V microstructures are generated in Neper using a two-level tessellation strategy, with equiaxed prior-\(\beta\) grains constructed at the first level and the constituent \(\alpha\) and \(\beta\) grains introduced at the second level \citep{Quey_2022}. Each microstructure is embedded in a unit cubic domain and discretized using quadratic tetrahedral elements. A normalized characteristic length of 0.08 relative to the unit-domain length and a minimum mesh quality criterion of 1.0 are adopted. The prior-\(\beta\) crystallographic orientations are sampled from the BCC fundamental region, and the corresponding \(\alpha\)-grain orientations are subsequently generated by enforcing the BOR between the BCC \(\beta\) and HCP \(\alpha\) phases \citep{Niessen:nb5309}. As illustrated in Fig. \ref{fig:FEASetup} b), the mechanical response of each microstructure is computed using an elastic-viscoplastic crystal plasticity formulation implemented in FEPX \citep{Quey_2022}. Uniaxial tensile loading is applied at a constant strain rate of \(0.001\,\mathrm{s^{-1}}\) up to a total strain of 1.5\%. The resulting reaction forces are converted into homogenized engineering stresses using the initial cross-sectional area. As shown in Fig. \ref{fig:FEASetup} c),  the stress values at six discrete strain levels, 0.25\%,\;0.50\%,\;0.75\%,\;1.00\%,\;1.25\%,\;1.50\%, are used as supervised labels for forward stress prediction. The derived database properties are computed directly from these discrete stress labels. Specifically, the elastic modulus is represented by the secant modulus at 0.25\% strain. Among the available discrete stress labels, the stress at 1.00\% strain lies closest to the yield region indicated by the 0.2\%-offset convention. It is therefore used as a proxy for yield stress and is referred to as the yield-stress proxy throughout this work. This definition provides a consistent discrete label without requiring interpolation of the simulated stress-strain curves.

The property statistics of the Ti-6Al-4V database are summarized in Table~\ref{tab:database_exp_range} with experimentally reported ranges of elastic modulus and yield stress also included for comparison \cite{met9040447,met11010104}. The mean values of elastic modulus and yield-stress proxy derived from the database fall within the experimental range, showing good agreement between the simulated database labels and reported Ti-6Al-4V mechanical properties.

\begin{table}[htbp]
  \centering
  \caption{Ti-6Al-4V database property statistics and experimental property ranges.}
    \begin{tabular}{lrrrrl}
    \toprule
          & \multicolumn{1}{l}{Mean} & \multicolumn{1}{l}{SD} & \multicolumn{1}{l}{Min.} & \multicolumn{1}{l}{Max.} & Exp. \cite{ met9040447, met11010104} \\
    \midrule
    $\alpha $ volume fraction, -& 0.54  & 0.15  & 0.24& 0.98&  \\
    Elastic modulus, GPa& 111.91 & 5.81  & 93.91& 130.94& 105.31-117.00 \\
    Yield stress, GPa& 0.94  & 0.03  & 0.82& 1.09& 0.77-1.19 \\
    \bottomrule
    \end{tabular}
  \label{tab:database_exp_range}
\end{table}

\section{Grain GNN for Forward Property Prediction}

\begin{figure}[ht]
    \begin{center}
     \centering
     \includegraphics[width=450pt]{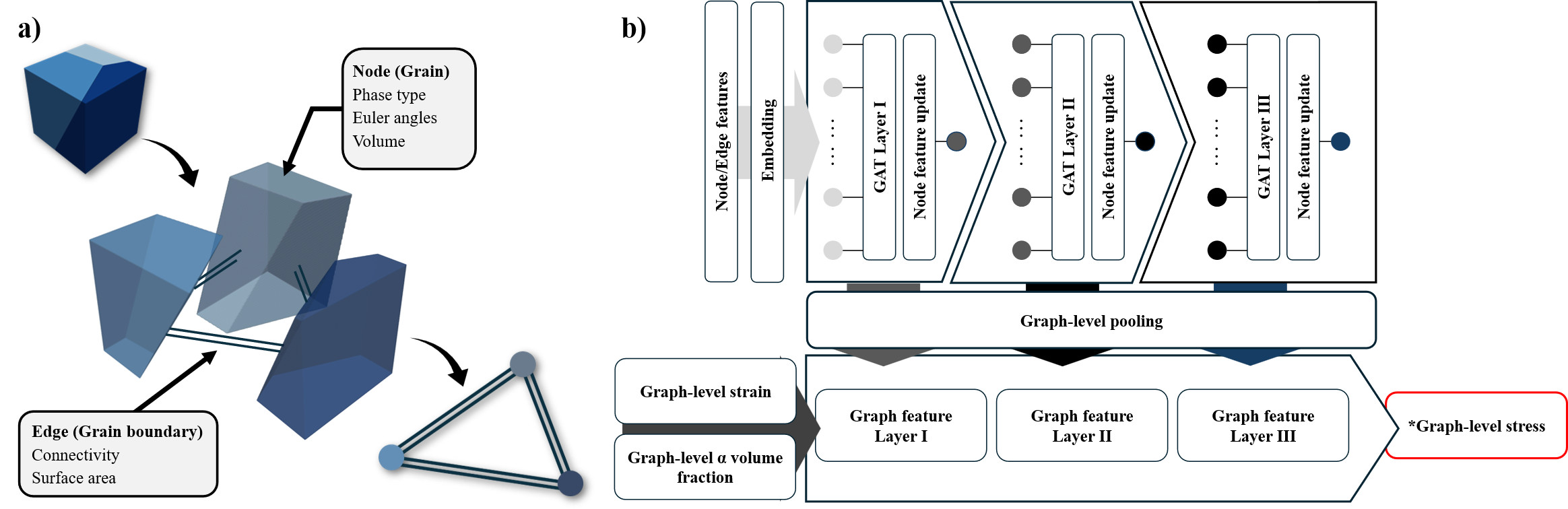}
    \end{center}
    \vspace{-1em}
    \caption{Grain graph representation and forward grain GNN architecture: a) Directed grain graph with nodes corresponding to grains and edges corresponding to grain boundaries. b) Schematic of the forward GNN architecture.}
    \label{fig:grainGNNForward}
    \end{figure}

The Ti-6Al-4V microstructure is represented using a grain graph for rapid prediction of its mechanical response. As illustrated in Fig. \ref{fig:grainGNNForward} a), each microstructure is converted into a directed graph, where each node corresponds to a crystalline grain and each edge corresponds to a grain boundary shared by neighboring grains. Each physical grain boundary is represented by two directed edges to allow message passing in both directions. The graph descriptors include grain phase, crystallographic orientation, grain volume, grain boundary connectivity, and grain boundary surface area. In the GNN model, the geometric descriptors are used in normalized form for better generalizability: grain volume is represented as grain volume fraction, and grain boundary surface area is represented as contact area fraction. The node features include the binary phase label, the crystallographic orientation, and the grain volume fraction. The crystallographic orientation is described by the Euler angles and encoded using sine-cosine functions \citep{bunge1982}. Grain boundary connectivity is encoded by the graph adjacency, while the active edge attribute is the contact area fraction, which provides local grain boundary geometry information beyond connectivity alone.

The forward model follows a graph attention architecture related to the GAT-based framework of Gao et al.~\cite{GAO2025117844}, with several modifications to better use the grain and grain boundary descriptors. As demonstrated in Fig. \ref{fig:grainGNNForward} b), the node and edge features are first passed through multilayer perceptron (MLP) embedding layers before being processed by GAT message passing layers \cite{brody2022gatv2}. After each GAT layer, graph-level descriptors are extracted using multi-statistic pooling. The pooled graph descriptors are then concatenated with the graph-level \(\alpha\)-phase volume fraction and strain condition before the final MLP regression head. The model is trained to predict the stress response at six prescribed strain levels: 0.25\%, 0.50\%, 0.75\%, 1.00\%, 1.25\%, and 1.50\%. The dataset is split at the graph level using a 4,141/518/518 (train/validation/test) split, ensuring that microstructures in the validation and test sets are not seen during training.

For a grain graph \(G_i=(V_i,\mathcal{E}_i)\), nodes represent grains and directed edges represent grain boundary contacts. The forward grain GNN predicts the stress at strain level \(\varepsilon_k\) as
\begin{equation}
\hat{\sigma}_{ik} = f_{\theta}(G_i,\varepsilon_k).
\end{equation}
For node \(u\in V_i\), the input feature vector is defined as
\begin{equation}
\mathbf{x}_u=
\left[
 z_u
 ;
 f_{\mathrm{enc}}(\phi_{1,u},\Phi_u,\phi_{2,u})
 ;
 f_u
\right],
\end{equation}
where \(z_u\in\{0,1\}\) is the binary phase indicator, with \(1\) for \(\alpha\) and \(0\) for \(\beta\), and \(f_u=v_u/v_{\mathrm{total}}\) is the volume fraction of grain \(u\). The variables \(\phi_{1,u},\Phi_u,\phi_{2,u}\) are the Bunge Euler angles of grain \(u\) \citep{bunge1982}. To reduce discontinuities associated with angular periodicity, the Euler angles are encoded using a sine-cosine representation:
\begin{equation}
f_{\mathrm{enc}}(\phi_{1,u},\Phi_u,\phi_{2,u})
=
\left[
\sin\phi_{1,u},\;
\sin\Phi_u,\;
\sin\phi_{2,u},\;
\cos\phi_{1,u},\;
\cos\Phi_u,\;
\cos\phi_{2,u}
\right].
\end{equation}
A directed edge \((v,u)\in \mathcal{E}_i\) represents the grain boundary between grain \(u\) and its neighboring grain \(v\). The edge feature is defined as
\begin{equation}
\mathbf{e}_{vu}=\left[f_{vu}^{A}\right],
\end{equation}
where \(f_{vu}^{A}\) is the fraction of the surface area of grain \(v\) that is in contact with grain \(u\). The graph-level \(\alpha\)-phase volume fraction is defined as
\begin{equation}
g_i = v_i^{\alpha} = \sum_{u\in V_i} f_u z_u .
\end{equation}
Before message passing, node-, edge-, and graph-level features, strain conditions, and stress labels are normalized independently using statistics computed from the training set. For each graph-strain pair
\((G_i,\varepsilon_k)\), the normalized target stress is defined as
\begin{equation}
\tilde{\sigma}_{ik}
=
\frac{\sigma_{ik}-\mu_{\sigma,k}}{s_{\sigma,k}},
\end{equation}
where \(\mu_{\sigma,k}\) and \(s_{\sigma,k}\) are the training-set mean and standard deviation of the stress at strain level \(\varepsilon_k\), respectively. The normalized node and edge features are then mapped through learnable embedding transformations:
\begin{equation}
\mathbf{h}_u^{(0)}
=
\phi_{\mathrm{node}}(\tilde{\mathbf{x}}_u),
\qquad
\boldsymbol{\eta}_{vu}
=
\phi_{\mathrm{edge}}(\tilde{\mathbf{e}}_{vu}).
\end{equation}
The grain GNN is implemented using PyTorch Geometric (PyG) \citep{fey2019fastgraphrepresentationlearning}. The message passing backbone consists of three GATv2 layers with a hidden dimension of 128 and attention heads (3,3,2) \citep{brody2022gatv2}. For layer \(\ell\), the attention-weighted message to node \(u\) is computed as

\begin{equation}
\mathbf{m}_u^{(\ell)}
=
\sum_{v\in\mathcal{N}(u)}
\alpha_{vu}^{(\ell)}
\mathbf{W}_{t}^{(\ell)}\mathbf{h}_v^{(\ell)},
\end{equation}
where the attention coefficient for the directed edge \((v,u)\) is
\begin{equation}
\alpha_{vu}^{(\ell)}
=
\frac{
\exp\!\left(
\mathbf{a}^{(\ell)\top}
\mathrm{LeakyReLU}
\left(
\mathbf{W}_{s}^{(\ell)}\mathbf{h}_u^{(\ell)}
+
\mathbf{W}_{t}^{(\ell)}\mathbf{h}_v^{(\ell)}
+
\mathbf{W}_{e}^{(\ell)}\boldsymbol{\eta}_{vu}
\right)
\right)
}{
\sum_{q\in\mathcal{N}(u)}
\exp\!\left(
\mathbf{a}^{(\ell)\top}
\mathrm{LeakyReLU}
\left(
\mathbf{W}_{s}^{(\ell)}\mathbf{h}_u^{(\ell)}
+
\mathbf{W}_{t}^{(\ell)}\mathbf{h}_q^{(\ell)}
+
\mathbf{W}_{e}^{(\ell)}\boldsymbol{\eta}_{qu}
\right)
\right)
},
\end{equation}
where \(\mathcal{N}(u)\) denotes the neighboring grains connected to grain \(u\). After the attention update, the node state is updated through residual addition, layer normalization \(\mathrm{LN}(\cdot)\) \citep{ba2016layernorm}, dropout with \(p_{\mathrm{drop}}=0.10\) \citep{srivastava2014dropout}, and a feed-forward MLP:
\begin{equation}
\bar{\mathbf{h}}_u^{(\ell+1)}
=
\mathrm{LN}\!\left(
\mathbf{h}_u^{(\ell)}
+
\mathrm{Dropout}_{p_{\mathrm{drop}}}\!\left(\mathbf{m}_u^{(\ell)}\right)
\right),
\end{equation}
\begin{equation}
\mathbf{h}_u^{(\ell+1)}
=
\mathrm{LN}\!\left(
\bar{\mathbf{h}}_u^{(\ell+1)}
+
\mathrm{Dropout}_{p_{\mathrm{drop}}}\!\left(
\phi_{\mathrm{ffn}}^{(\ell)}
\left(
\bar{\mathbf{h}}_u^{(\ell+1)}
\right)
\right)
\right).
\end{equation}
After each message passing layer, graph-level pooling is performed using the grain volume fraction weights \(w_u=f_u\), with \(\sum_{u\in V_i} w_u=1\). The weighted mean, weighted standard deviation, and max pooling are computed as
\begin{equation}
\boldsymbol{\mu}_i^{(\ell)}
=
\sum_{u \in V_i} w_u \, \mathbf{h}_u^{(\ell)},
\end{equation}
\begin{equation}
\boldsymbol{\sigma}_i^{(\ell)}
=
\sqrt{
\sum_{u \in V_i} w_u
\left(\mathbf{h}_u^{(\ell)}-\boldsymbol{\mu}_i^{(\ell)}\right)^2
},
\end{equation}
\begin{equation}
\mathbf{m}_i^{(\ell)}
=
\max_{u \in V_i} \mathbf{h}_u^{(\ell)} .
\end{equation}
The pooled graph descriptor is constructed as
\begin{equation}
\mathbf{p}_i^{(\ell)}
=
\left[
\boldsymbol{\mu}_i^{(\ell)};
\boldsymbol{\sigma}_i^{(\ell)};
\mathbf{m}_i^{(\ell)}
\right].
\end{equation}
The final graph-level vector is obtained by concatenating the pooled descriptors from the three GATv2 layers with the normalized graph-level \(\alpha\)-phase volume fraction and the normalized strain condition:
\begin{equation}
\mathbf{z}_{ik}
=
\left[
\mathbf{p}_i^{(1)};
\mathbf{p}_i^{(2)};
\mathbf{p}_i^{(3)};
\tilde{g}_i;
\tilde{\varepsilon}_k
\right].
\end{equation}
Finally, the normalized stress is predicted by a MLP head,
\begin{equation}
\hat{\tilde{\sigma}}_{ik}
=
\phi_{\mathrm{head}}(\mathbf{z}_{ik}),
\end{equation}
SiLU activation \citep{DBLP:journals/corr/ElfwingUD17} is used in the GAT forward blocks and the hidden layers of the prediction head. The stress is mapped back to the physical stress scale as
\begin{equation}
\hat{\sigma}_{ik}
=
\mu_{\sigma,k}
+
s_{\sigma,k}\hat{\tilde{\sigma}}_{ik}.
\end{equation}
The forward predictor is trained using the mean squared error between the predicted and true normalized stresses:
\begin{equation}
\mathcal{L}_{\mathrm{pred}}
=
\frac{1}{|\mathcal{D}|}
\sum_{(i,k)\in\mathcal{D}}
\left\|
\hat{\tilde{\sigma}}_{ik}
-
\tilde{\sigma}_{ik}
\right\|^2,
\end{equation}
where \(\mathcal{D}\) denotes the graph-strain pairs in the training set. The training is performed using the AdamW optimizer with an initial learning rate of 0.0001 and weight decay of 0.0001 \cite{loshchilov2019decoupledweightdecayregularization}. A batch size of 64 is used, and training is conducted for a maximum of 500 epochs, with early stopping applied after 80 consecutive epochs without improvement in the validation loss. 

\begin{figure}[ht]
    \begin{center}
     \centering
     \includegraphics[width=400pt]{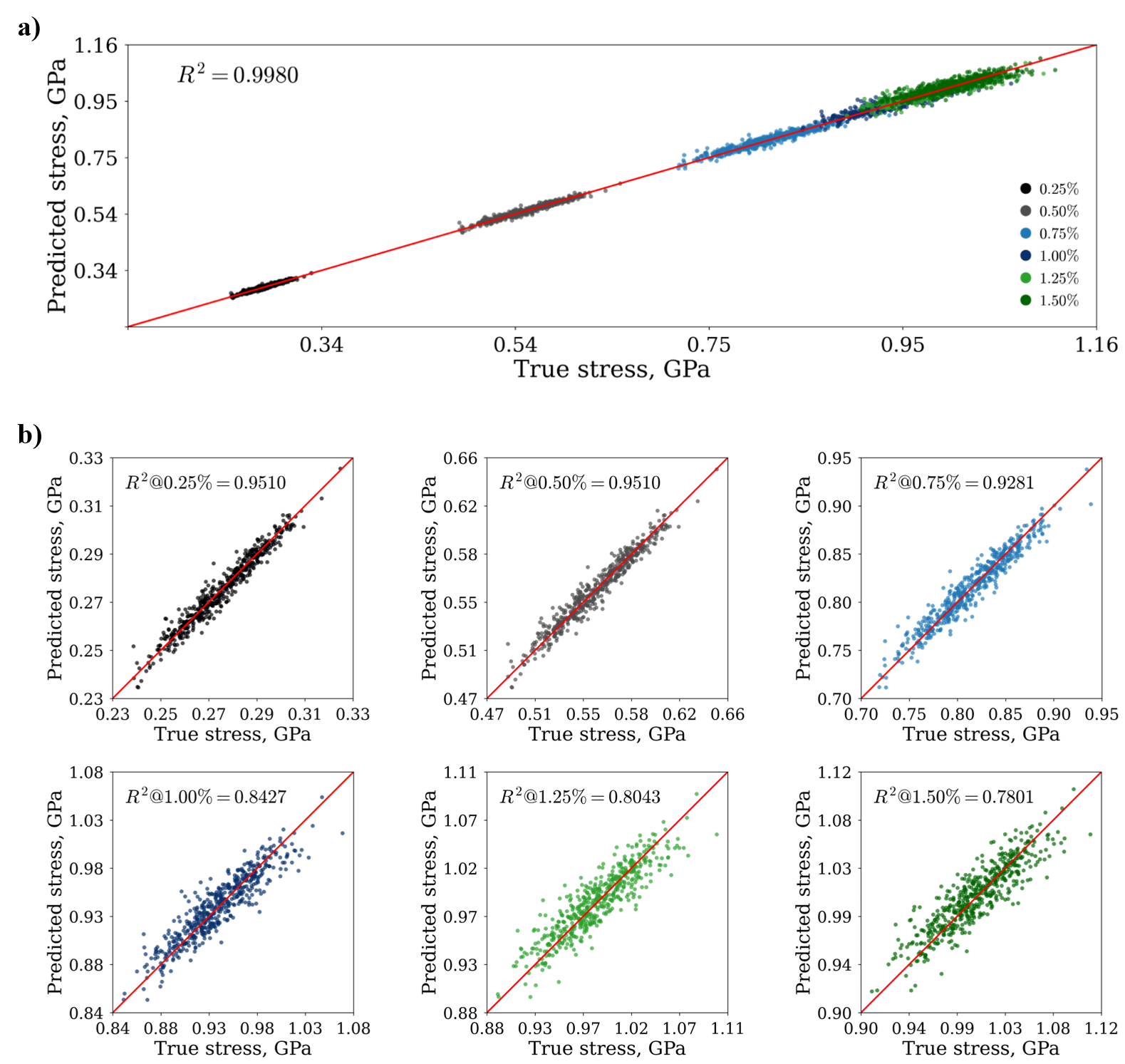}
    \end{center}
    \vspace{-1em}
    \caption{Forward stress prediction performance of the grain GNN presented by the correlation between FEA computed stresses and GNN predicted stresses: a) across all strain levels; b) at each prescribed strain level.}
    \label{fig:GNNTrainingCorrelation}
    \end{figure}

The forward prediction performance is evaluated using the test set. Fig. \ref{fig:GNNTrainingCorrelation} a) shows the correlation between FEA computed stresses and GNN predicted stresses across all strain levels, while Fig. \ref{fig:GNNTrainingCorrelation} b) provides the corresponding true versus predicted stress correlations at individual strain levels. Overall, the grain GNN achieves accurate stress prediction, with an overall \(R^2\) of 0.998 and mean absolute error (MAE) of 0.008~GPa. The results from individual strain levels show that the prediction performance is strongest in the elastic regime and decreases at higher strain levels. This trend is consistent with the increasingly nonlinear mechanical response after the onset of plasticity, where the microstructure-property mapping becomes more difficult to resolve using a surrogate model.

\begin{figure}[ht]
    \begin{center}
     \centering
     \includegraphics[width=400pt]{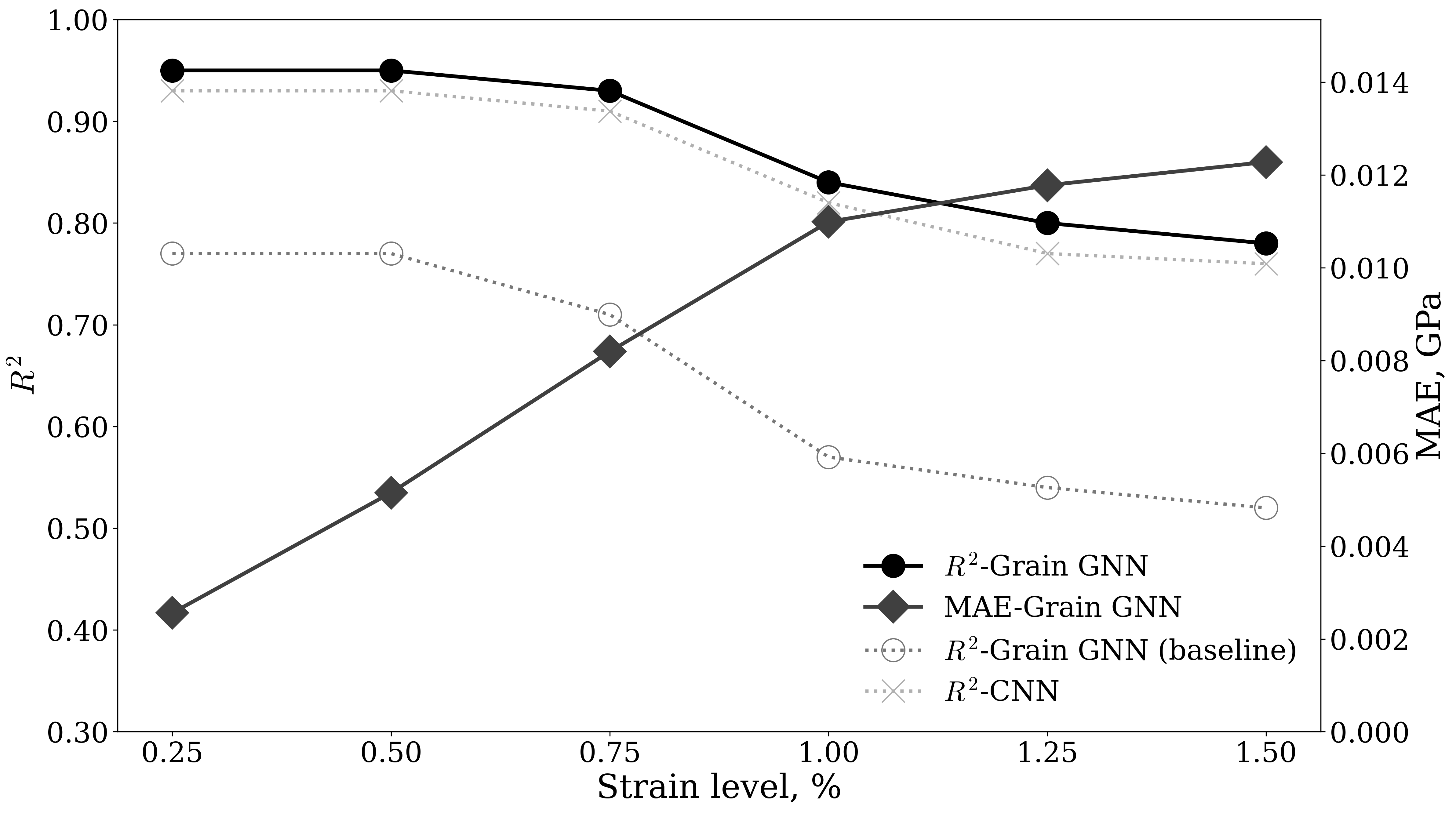}
    \end{center}
    \vspace{-1em}
    \caption{\(R^2\) and MAE of grain GNN at each strain level, and \(R^2\) of grain GNN baseline and CNN model.}
    \label{fig:GNNTrainingSummary}
    \end{figure}

The variation of \(R^2\) and MAE with strain is summarized in Fig. \ref{fig:GNNTrainingSummary}. The MAE increases rapidly from 0.25\% strain to approximately 1.0\% strain, after which it begins to plateau. This behavior is associated with the changing magnitude of stress values across the strain range. From 0.25\% to 1.0\% strain, the stress increases rapidly, so a comparable relative prediction error produces a larger absolute error and leads to a sharp increase in MAE. At higher strain levels, the change of stress response becomes less steep, and the increase in MAE becomes less pronounced. The \(R^2\) score decreases sharply between approximately 0.75\% and 1.00\% strain, corresponding to the transition from predominantly elastic deformation to plastic deformation. This indicates that the relative ability of the model to resolve differences between the graphs in stress response becomes more limited in the plastic regime.

For comparison, the \(R^2\) values of the grain GNN are benchmarked against a baseline grain GNN model adapted from Gao et al.~\cite{GAO2025117844} and the voxel-based CNN model reported by Shang et al.~\cite{SHANG202371}. The voxel-based CNN achieves comparable \(R^2\) values at the reported strain levels, but it operates on a voxelized 3D microstructure representation. In contrast, the grain GNN achieves competitive accuracy using a compact graph representation. Specifically, for the representative microstructure with 38 grains shown in Fig. \ref{fig:FEASetup} a), a \(32^3\) voxelized representation contains 32,768 voxel locations before accounting for feature channels. In comparison, the graph representation encodes the same grain microstructure using only 38 grain nodes and 131 undirected grain boundary edges. The baseline grain GNN model shows a lower accuracy, indicating that the enhanced graph representation and GNN architecture, including grain boundary properties as edge features, node and edge feature embeddings, and multi-statistic graph pooling, improve the forward stress prediction capability.

\section{Conditional Microstructure Generation and Finite Element Validation}
\begin{figure}[ht]
    \begin{center}
     \centering
     \includegraphics[width=400pt]{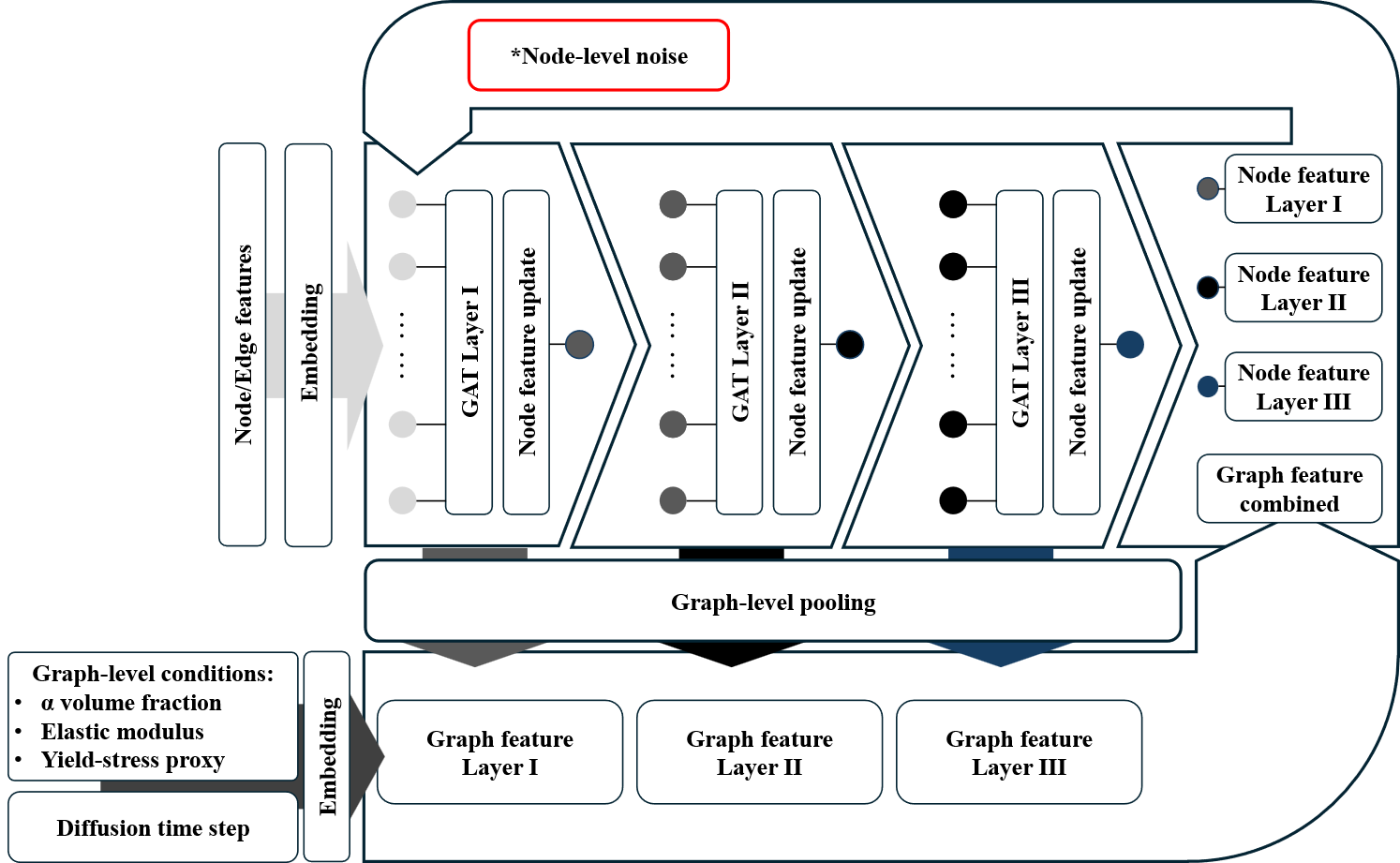}
    \end{center}
    \vspace{-1em}
    \caption{Schematic of the conditional graph denoising network used in the diffusion model.}
    \label{fig:GNNDiffusion}
    \end{figure}

The inverse design model uses the same GAT message passing backbone as the grain GNN for forward property prediction. Instead of predicting graph-level stress values, the diffusion GNN predicts the node-level noise added to the grain-level design variables, which consists of the phase and crystallographic orientation of each grain. As illustrated in Fig. \ref{fig:GNNDiffusion}, the graph-level features are combined with the target property conditions and the diffusion timestep embedding to form a graph-level feature vector. This graph-level feature vector is then broadcast back to the nodes and fused with the node-level features, allowing the model to predict node-level noise using both node-level grain information and graph-level design target information.

The binary phase is represented by a continuous phase logit \(\ell_{u,0}\in\mathbb{R}\), with associated phase probability and binary \(\alpha\)-phase assignment
\begin{equation}
p_{u,0}
=
\sigma(\ell_{u,0}),
\qquad
z_{u,0}
=
\mathbb{I}[p_{u,0}>0.5].
\end{equation}
The Euler angles \((\phi_{1,u},\Phi_u,\phi_{2,u}) \in [0,2\pi)\times[0,\pi]\times[0,2\pi)\) are mapped to a rotation matrix \(\mathbf{R}_{u,0}\in SO(3)\) and then to a rotation vector using the logarithmic map, \(\mathrm{Log}(\cdot)\) \citep{gallego2015compact,Rowenhorst_2015,zhou2019continuity,jagvaral2023so3}:
\begin{equation}
\boldsymbol{\omega}_{u,0}
=
\mathrm{Log}(\mathbf{R}_{u,0})
\in\mathbb{R}^{3}.
\end{equation}
The node-level diffusion state is therefore
\begin{equation}
\mathbf{y}_{u,0}
=
\left[
\boldsymbol{\omega}_{u,0};
\ell_{u,0}
\right]
\in\mathbb{R}^{4},
\end{equation}
and the graph-level state is denoted by
\begin{equation}
\mathbf{Y}_0=\{\mathbf{y}_{u,0}\}_{u\in V_i}.
\end{equation}
Following DDPMs \citep{ho2020ddpm}, Gaussian noise is injected into the design variables through the forward noising process
\begin{equation}
q(\mathbf{Y}_t\mid \mathbf{Y}_0)
=
\mathcal{N}
\!\left(
\sqrt{\bar{\alpha}_t}\,\mathbf{Y}_0,
(1-\bar{\alpha}_t)\mathbf{I}
\right),
\end{equation}
with
\begin{equation}
\alpha_t = 1-\beta_t,
\qquad
\bar{\alpha}_t = \prod_{s=1}^{t}\alpha_s,
\end{equation}
where \(\beta_t\) follows a cosine schedule \citep{nichol2021improved}. At diffusion step \(t\), the node feature vector is rebuilt from the current diffusion state as
\begin{equation}
\mathbf{x}^{\mathrm{diff}}_{u,t}
=
\left[
\boldsymbol{\omega}_{u,t};
\ell_{u,t};
f_u
\right],
\end{equation}
where \(f_u\) is the grain volume fraction. The edge feature retains the same geometric form used in the forward model:
\begin{equation}
\mathbf{e}^{\mathrm{diff}}_{vu,t}
=
\left[
f_{vu}^{A}
\right],
\end{equation}
where \(f_{vu}^{A}\) is the directed contact area fraction. To preserve consistency with the forward model, the target elastic modulus and yield-stress proxy are translated to stress space conditioning targets as $\sigma_{0.25}^{*}=E^{*}\varepsilon_{\mathrm{el}}$ and $\sigma_{1.00}^{*}=Y^{*}$, where \(\varepsilon_{\mathrm{el}}=0.0025\). The graph-level conditioning target is defined as
\begin{equation}
\mathbf{q}_i^{*}
=
\left[
\sigma_{0.25}^{*}
;
\sigma_{1.00}^{*}
;
v_f^{\alpha *}
\right],
\end{equation}
where \(v_f^{\alpha *}\) is the target \(\alpha\)-phase volume fraction. Each conditioning component is normalized using statistics computed over the diffusion model dataset, and the resulting normalized property target, \(\tilde{\mathbf{q}}_i^{*}\), is embedded using a MLP:
\begin{equation}
\mathbf{c}_i = \phi_c(\tilde{\mathbf{q}}_i^{*}).
\end{equation}

The reverse diffusion denoiser uses the same GATv2 block architecture and multi-statistic graph-level pooling strategy as the forward GNN. However, the denoiser has separate node and edge input embeddings, separate trainable parameters, and a different output head for node-level noise prediction. The diffusion node and edge features are first embedded as
\begin{equation}
\mathbf{h}_{u,t}^{(0),\mathrm{diff}}
=
\phi_{\mathrm{node}}^{\mathrm{diff}}
\left(
\mathbf{x}^{\mathrm{diff}}_{u,t}
\right),
\qquad
\boldsymbol{\eta}_{vu}^{\mathrm{diff}}
=
\phi_{\mathrm{edge}}^{\mathrm{diff}}
\left(
\mathbf{e}^{\mathrm{diff}}_{vu,t}
\right).
\end{equation}
The embedded graph is then processed by three GATv2 message passing layers in the same manner as the forward model:
\begin{equation}
\mathbf{h}_{u,t}^{(\ell),\mathrm{diff}}
=
\mathrm{GATv2}^{(\ell),\mathrm{diff}}
\left(
\mathbf{h}_{u,t}^{(\ell-1),\mathrm{diff}},
\boldsymbol{\eta}_{vu}^{\mathrm{diff}},
\mathcal{E}_i 
\right),
\qquad
\ell=1,2,3.
\end{equation}
After each layer, the same pooling operator used in the forward model is applied:
\begin{equation}
\mathbf{p}^{(\ell),\mathrm{diff}}_{i,t}
=
\left[
\boldsymbol{\mu}^{(\ell),\mathrm{diff}}_{i,t}
;
\boldsymbol{\sigma}^{(\ell),\mathrm{diff}}_{i,t}
;
\mathbf{m}^{(\ell),\mathrm{diff}}_{i,t}
\right],
\end{equation}
where the weighted mean and standard deviation use the grain volume-fraction weights \(w_u=f_u\).

The diffusion timestep is encoded using a sinusoidal positional embedding, \(\mathrm{PE}(\cdot)\), and then mapped through a MLP \citep{nichol2021improved}:
\begin{equation}
\mathbf{t}_t
=
\phi_t
\left(
\mathrm{PE}(t)
\right).
\end{equation}
The graph-level context vector is then formed by combining the diffusion pooled descriptors, timestep embedding, and conditioning vector:
\begin{equation}
\mathbf{g}_{i,t}
=
\phi_g
\left[
\mathbf{p}^{(1),\mathrm{diff}}_{i,t}
;
\mathbf{p}^{(2),\mathrm{diff}}_{i,t}
;
\mathbf{p}^{(3),\mathrm{diff}}_{i,t}
;
\mathbf{t}_t
;
\mathbf{c}_i
\right].
\end{equation}
The graph-level context is broadcast back to the node level and used to predict the additive diffusion noise for each node state:
\begin{equation}
\hat{\boldsymbol{\epsilon}}_{u,t}
=
\phi_{\mathrm{out}}
\left[
\mathbf{h}^{(1),\mathrm{diff}}_{u,t}
;
\mathbf{h}^{(2),\mathrm{diff}}_{u,t}
;
\mathbf{h}^{(3),\mathrm{diff}}_{u,t}
;
\mathbf{g}_{i,t}
\right],
\qquad
\hat{\boldsymbol{\epsilon}}_{u,t}\in\mathbb{R}^{4}.
\end{equation}
The denoiser is trained with the standard DDPM noise prediction objective \citep{ho2020ddpm}:
\begin{equation}
\mathcal{L}_{\mathrm{diff}}
=
\mathbb{E}_{i,t,\boldsymbol{\epsilon}}
\left[
\left\|
\boldsymbol{\epsilon}
-
\hat{\boldsymbol{\epsilon}}_{\theta}
(\mathbf{Y}_t,t,\mathbf{c}_i)
\right\|^2
\right].
\end{equation}

Training is performed using the AdamW optimizer with a learning rate of 0.0003, a weight decay of 0.0001, and a batch size of 64 \cite{loshchilov2019decoupledweightdecayregularization}. The dataset is divided into 4,659 training and 518 validation graphs. Training is limited to 500 epochs, with early stopping applied after 80 consecutive epochs without improvement in the validation loss. During training, the conditioning vector is randomly dropped with probability \(p_{\mathrm{cond\text{-}drop}}=0.10\) to enable classifier-free guidance (CFG) during sampling \citep{hosalimans2022cfg}:
\begin{equation}
\hat{\boldsymbol{\epsilon}}_{\theta}^{\mathrm{cfg}}
=
\hat{\boldsymbol{\epsilon}}_{\theta}^{\varnothing}
+
s\left(
\hat{\boldsymbol{\epsilon}}_{\theta}^{\mathrm{cond}}
-
\hat{\boldsymbol{\epsilon}}_{\theta}^{\varnothing}
\right),
\end{equation}
where \(\hat{\boldsymbol{\epsilon}}_{\theta}^{\varnothing}\) is the unconditional noise prediction, \(\hat{\boldsymbol{\epsilon}}_{\theta}^{\mathrm{cond}}\) is the conditional noise prediction, and \(s\) is the guidance scale which is defined as \(s=6\) in this work. 

The reverse diffusion update follows the standard DDPM Gaussian posterior parameterization \citep{ho2020ddpm,nichol2021improved}:
\begin{equation}
\boldsymbol{\mu}_{\theta}
(\mathbf{Y}_t,t,\mathbf{c}_i)
=
\frac{1}{\sqrt{\alpha_t}}
\left(
\mathbf{Y}_t
-
\frac{\beta_t}{\sqrt{1-\bar{\alpha}_t}}
\hat{\boldsymbol{\epsilon}}_{\theta}^{\mathrm{cfg}}
(\mathbf{Y}_t,t,\mathbf{c}_i)
\right),
\end{equation}
with posterior variance
\begin{equation}
\tilde{\beta}_t
=
\beta_t
\frac{1-\bar{\alpha}_{t-1}}{1-\bar{\alpha}_t}.
\end{equation}
The sampling step is performed as
\begin{equation}
\mathbf{Y}_{t-1}^{\mathrm{raw}}
=
\boldsymbol{\mu}_{\theta}
(\mathbf{Y}_t,t,\mathbf{c}_i)
+
\mathbb{I}[t>1]\sqrt{\tilde{\beta}_t}\,\boldsymbol{\xi},
\qquad
\boldsymbol{\xi}\sim\mathcal{N}(\mathbf{0},\mathbf{I}).
\end{equation}

After each reverse step, a graph-level phase fraction correction is applied in logit space. For the raw logits \(\ell_{u,t-1}^{\mathrm{raw}}\), a scalar shift \(\Delta_{t-1}\) is computed such that
\begin{equation}
\sum_{u\in V_i}
f_u
\sigma(\ell_{u,t-1}^{\mathrm{raw}}+\Delta_{t-1})
=
v_f^{\alpha *}.
\end{equation}
The corrected phase logits are then
\begin{equation}
\ell_{u,t-1}^{\mathrm{corr}}
=
\ell_{u,t-1}^{\mathrm{raw}}
+
\Delta_{t-1}.
\end{equation}
The corrected logits \(\ell_{u,t-1}^{\mathrm{corr}}\) replace the raw logits \(\ell_{u,t-1}^{\mathrm{raw}}\) before the next reverse step. After the reverse diffusion process, the output phase probability and binary phase assignment are recovered as
\begin{equation}
p_u^{\mathrm{out}}
=
\sigma(\ell_u^{\mathrm{corr}}),
\qquad
z_u^{\mathrm{out}}
=
\mathbb{I}[p_u^{\mathrm{out}}>0.5].
\end{equation}
The crystallographic orientation is recovered from the output rotation vector using the exponential mapping, \(\mathrm{Exp}(\cdot)\), and converted back to Euler angles, \(\mathrm{EA}(\cdot)\) \citep{gallego2015compact,Rowenhorst_2015}:
\begin{equation}
\mathbf{R}_u^{\mathrm{out}}
=
\mathrm{Exp}(\boldsymbol{\omega}_u^{\mathrm{out}}),
\qquad
(\phi_{1,u}^{\mathrm{out}},\Phi_u^{\mathrm{out}},\phi_{2,u}^{\mathrm{out}})
=
\mathrm{EA}(\mathbf{R}_u^{\mathrm{out}}).
\end{equation}

\begin{figure}[ht]
    \begin{center}
     \centering
     \includegraphics[width=450pt]{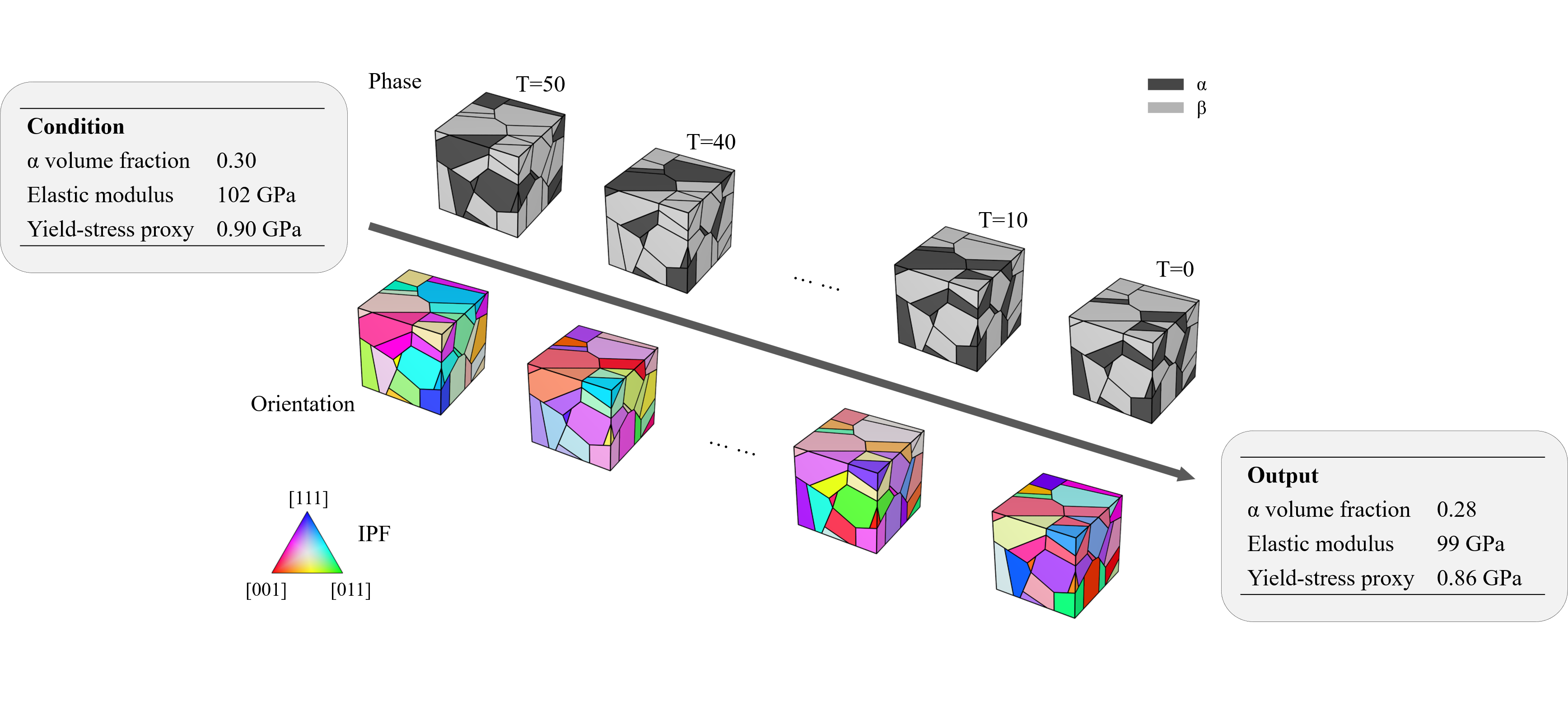}
    \end{center}
    \vspace{-1em}
    \caption{Illustration of the conditional reverse diffusion process for grain microstructure inverse design.}
    \label{fig:DiffusionIllustration}
    \end{figure}

As illustrated in Fig. \ref{fig:DiffusionIllustration}, the conditional sampling begins by applying the forward diffusion process to the phase and orientation variables of an input grain graph up to the selected starting timestep, \(T=50\). The reverse diffusion process then gradually maps the noised phase and orientation fields from \( T=50\) back to \(T=0\), generating a clean candidate microstructure conditioned on the prescribed property targets. 

\subsection{In-Envelope and Extrapolative Property Targets}

Two property target scenarios are considered to evaluate the conditional grain graph diffusion framework. The first is an in-envelope property-targeting task, for which the prescribed elastic modulus, yield-stress proxy, and $\alpha$-phase volume fraction are located within the property distribution represented by the reference database. This regime evaluates the ability of the framework to generate microstructures matching a feasible combination of prescribed properties. The second is an extrapolative high-property targeting task. In this case, the prescribed elastic modulus and yield-stress proxy extend beyond the corresponding upper bounds observed in the reference database, while a high $\alpha$-phase volume fraction is prescribed. It evaluates whether conditional generation can propose candidate microstructures outside the observed mechanical-property range. 

For each scenario, 50 grain microstructures are randomly selected from the database and used as input graphs for conditional generation. Two target cases are defined relative to the database statistics summarized in Table~\ref{tab:database_exp_range}. The design-to-spec case targets \(E^*=102\;\mathrm{GPa}\), \(Y^*=0.90\;\mathrm{GPa}\), and \(v_{\alpha}^*=0.3\), corresponding to the lower-tail region of the database distribution. The extreme-property case targets \(E^*=135\;\mathrm{GPa}\), \(Y^*=1.10\;\mathrm{GPa}\), and \(v_{\alpha}^*=0.75\), corresponding to the upper-tail region, with the mechanical targets extending beyond the database envelope. Specifically, \(E^*\) is 4.06~GPa above the database maximum of 130.94~GPa, while \(Y^*\) is 0.01~GPa above the database maximum of 1.09~GPa. Relative to the database statistics, these targets correspond to approximately 4.0 and 5.3 standard deviations above the mean, respectively. For each selected input graph, 32 candidate microstructure designs are generated and evaluated using the trained forward GNN model. The generated candidates are ranked using a target-matching score based on their deviations from the prescribed elastic modulus and yield-stress proxy targets:

\begin{equation}
S
=
w_E\frac{|E-E^*|}{|E^*|}
+
w_Y\frac{|Y-Y^*|}{|Y^*|} . 
\end{equation}
Here, \(E^*\) and \(Y^*\) denote the prescribed elastic modulus and yield-stress proxy targets, respectively. The weighting factors are set to \(w_E=w_Y=1\), giving equal weight to both target property terms.

    \begin{figure}[ht]
    \begin{center}
     \centering
     \includegraphics[width=450pt]{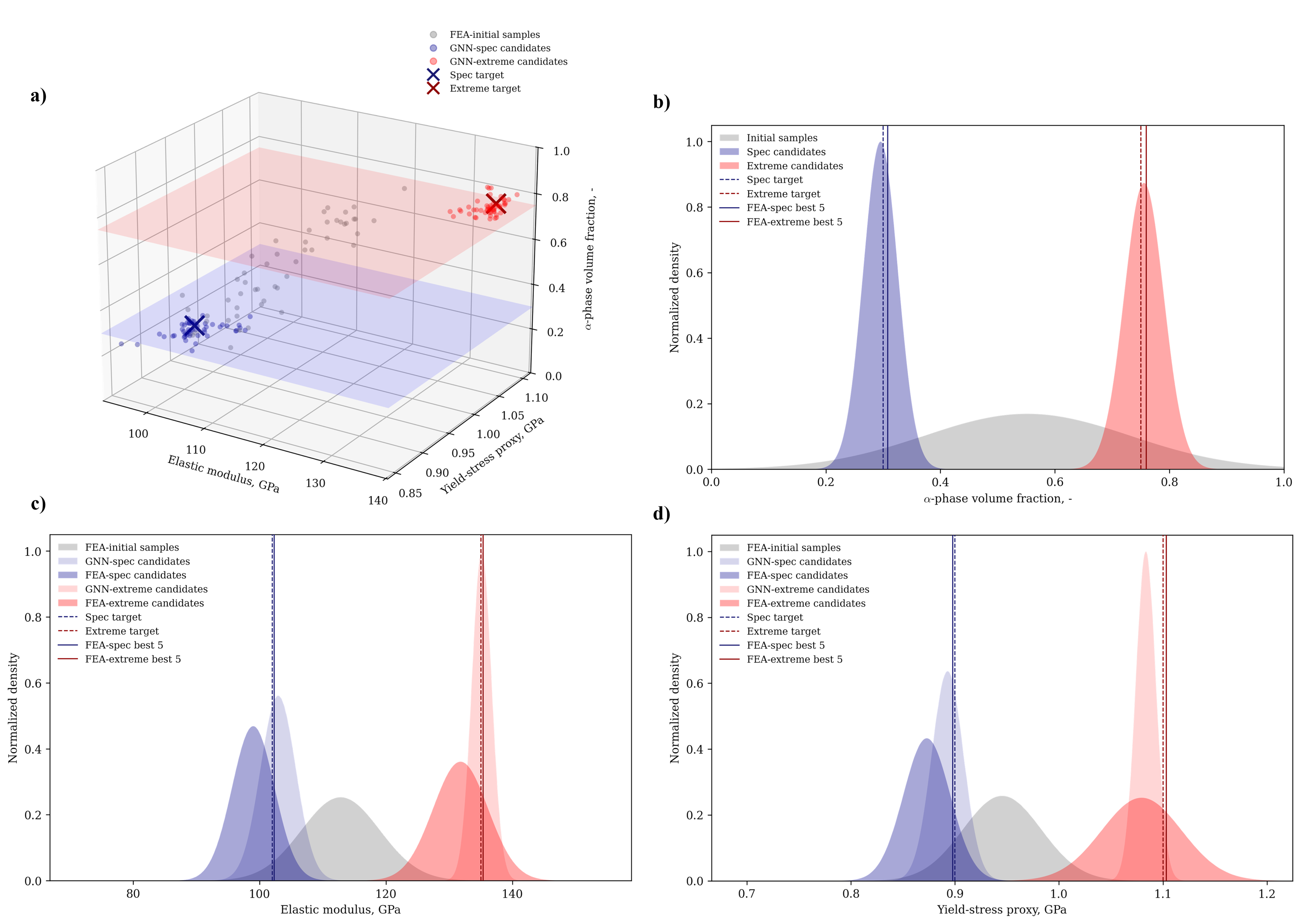}
    \end{center}
    \vspace{-1em}
    \caption{Property-guided microstructure generation for the design-to-spec and extreme-property design scenarios: a) Distribution of initial and generated candidate designs in the design property space. b-d) Distributions of \(\alpha\)-phase volume fraction, elastic modulus, and yield-stress proxy for the initial and generated candidate designs.}
    \label{fig:DiffusionDesign}
    \end{figure}

The properties of the initial candidate pool and the generated designs after the forward GNN screening are shown in Fig. \ref{fig:DiffusionDesign} a). The generated designs are shifted from the initial candidate pool and become clustered near the prescribed property targets in both the design-to-spec and extreme-property scenarios. For quantitative comparison, Fig. \ref{fig:DiffusionDesign} b-d) shows the distributions of \(\alpha\)-phase volume fraction, elastic modulus, and yield-stress proxy for the initial and generated candidate groups, while Table \ref{tab:conditioned_property_summary} summarizes the mean properties of the generated candidates from GNN and FEA, and the mean properties of the five best FEA-validated candidates. Values in parentheses denote the absolute relative error (ARE) computed with respect to the prescribed target value.

\begin{table}[htbp]
  \centering
  \caption{Property targeting performance of generated Ti-6Al-4V microstructure designs.}
  \tiny
  \setlength{\tabcolsep}{2pt}  
    \begin{tabular}{lccccccccc}
    \toprule
          &       & \multicolumn{2}{c}{\textbf{GNN}} & \multicolumn{2}{c}{\textbf{FEA}} &       & \multicolumn{3}{c}{\textbf{FEA Best 5}} \\
\cmidrule{1-6}\cmidrule{8-10}          & \textbf{\(\alpha\)-vf, -}& \textbf{E, GPa} & \textbf{Y, GPa}& \textbf{E, GPa} & \textbf{Y, GPa}&       & \textbf{\(\alpha\)-vf, -}& \textbf{E , GPa} & \textbf{Y, GPa}\\
\cmidrule{1-6}\cmidrule{8-10}    \textbf{Initial} & 0.552 & \multicolumn{2}{c}{-} & 112.82 & 0.945 &       & \multicolumn{3}{c}{-} \\
    \textbf{Spec target  } & 0.296 (1.48\%) & 102.95 (0.93\%) & 0.893 (0.78\%) & 99.00 (2.94\%) & 0.873 (3.00\%) &       & 0.308 (2.67\%) & 102.32 (0.31\%) & 0.898 (0.24\%) \\
    \textbf{Extreme target} & 0.755 (0.72\%) & 135.13 (0.10\%) & 1.084 (1.50\%) & 131.78 (2.38\%) & 1.079 (1.86\%) &       & 0.759 (1.20\%) & 135.37 (0.27\%) & 1.103 (0.29\%) \\
    \bottomrule
    \end{tabular}%
  \label{tab:conditioned_property_summary}%
\end{table}%

The generated \(\alpha\)-phase volume fractions align closely with the prescribed targets in both scenarios, with ARE of 1.48\% and 0.72\% respectively, confirming the effectiveness of the phase fraction control during reverse diffusion. In the design-to-spec case, the generated designs shift toward the prescribed low-modulus target while retaining agreement with the yield-stress proxy target, with an ARE of 0.93\% and 0.78\% for the elastic modulus and the yield-stress proxy, respectively. This case represents a constrained trade-off because elastic modulus and yield-stress proxy are strongly correlated in the database. Therefore, simultaneously lowering the elastic modulus while maintaining the prescribed yield-stress proxy target is challenging, and the generated distribution tends to fall between the two target requirements rather than collapsing exactly onto both.

In the extreme-property case, the generated designs show a strong shift toward the high-modulus and high-yield region. The mean elastic modulus coincides with the prescribed target well with an ARE of 0.10\%. The yield-stress proxy remains lower than the target on average, with an ARE of 1.50\%, suggesting that the prescribed yield-stress proxy target lies near or beyond the readily attainable range under the current phase and orientation design variables. Nevertheless, part of the generated distribution overlaps with the target yield-stress proxy region, indicating that viable high yield candidates can still be identified. Moreover, many of the generated extreme property designs exceed the maximum elastic modulus and yield-stress proxy observed in the selected input pool, despite being generated from only 50 selected input graphs from the full database of 5177.

For design validation, selected generated candidates are further evaluated using FEA, as shown in Fig. \ref{fig:DiffusionDesign} b-d). The FEA-computed elastic modulus and yield-stress proxy are generally slightly lower than the values predicted by the forward GNN. The scatter is more pronounced for yield-stress proxy than for elastic modulus, and the overall scatter is broader in the extreme property case than in the design-to-spec case. This is because both the reverse diffusion and candidate screening are based on the GNN surrogate, whereas FEA provides a higher-fidelity evaluation. The larger scatter in yield-stress proxy estimation is also consistent with the forward model results, where the grain GNN captures the elastic response at the lower strain more accurately than the yield stress at a higher strain. The discrepancy is more pronounced for the extreme-property target because this case requires extrapolation beyond the selected candidate property envelope, where the forward GNN prediction uncertainty is expected to be larger.

Nevertheless, although discrepancies remain between the GNN-predicted and FEA-computed properties, the FEA validation results still fall close to the prescribed target region. As shown in Table \ref{tab:conditioned_property_summary}, the average FEA-computed elastic modulus and yield-stress proxy of the best five generated candidates with the lowest selection score are in good agreement with the target values with a maximum ARE of 2.67\%. These results indicate that the conditional diffusion model can efficiently generate candidate grain microstructures close to the desired property targets, while FEA validation remains necessary to confirm the final candidate quality.

\subsection{Sensitivity Across Seeded Runs and Property Targets}

To examine the sensitivity of the conditional diffusion model to the selection of starting microstructures and to evaluate generation over a broader property range, two additional target sets were considered. The first, \(v_{\alpha}^*=0.54\), \(E^*=112~\mathrm{GPa}\), and \(Y^*=0.94~\mathrm{GPa}\), corresponds to the rounded mean properties of the present dataset. The second, \(v_{\alpha}^*=0.70\), \(E^*=118~\mathrm{GPa}\), and \(Y^*=0.97~\mathrm{GPa}\), represents an upper-tail target within the dataset property envelope. These targets complement the design-to-spec and extreme-property cases considered in the previous section. Each of the four target regimes was evaluated using three random-seed sample selections, with 50 starting microstructures selected for each run. Table \ref{tab:seed_target_sensitivity} reports the mean predicted properties across the three runs, together with the corresponding standard deviations. The AREs are also calculated for each target regime.

\begin{table}[htbp]
\scriptsize
  \centering
  \caption{Conditional diffusion results for different property targets with independently seeded runs.}
    \begin{tabular}{llllrrrr}
    \toprule
    \textbf{Target regime} & \multicolumn{3}{c}{\textbf{Mean}} &       & \multicolumn{3}{c}{\textbf{ARE}} \\
\cmidrule{2-4}\cmidrule{6-8}          & \textbf{\(\alpha\)-vf (SD), - }& \textbf{E (SD), GPa}& \textbf{Y (SD), GPa}&       & \multicolumn{1}{l}{\textbf{\(\alpha\)-vf}} & \multicolumn{1}{l}{\textbf{E}} & \multicolumn{1}{l}{\textbf{Y}} \\
    \midrule
    \textbf{Design-to-spec} & 0.298 (0.001) & 103.47 (0.31) & 0.897 (0.0021) 
&       & 0.65\%& 1.44\%& 0.30\%
\\
    \textbf{Dataset-mean}& 0.556 (0.006) & 110.79 (0.30) & 0.936 (0.0014) 
&       & 3.03\%& 1.08\%& 0.45\%
\\
    \textbf{Upper-tail}& 0.698 (0.007) & 118.05 (0.05) & 0.973 (0.0002) 
&       & 0.23\%& 0.04\%& 0.31\%
\\
    \textbf{Extreme-property}& 0.748 (0.003) & 134.69 (0.17) & 1.079 (0.0023) &       & 0.33\%& 0.23\%& 1.87\%\\
    \bottomrule
    \end{tabular}%
  \label{tab:seed_target_sensitivity}%
\end{table}%

As shown in Table \ref{tab:seed_target_sensitivity}, the candidate properties closely approach all four prescribed targets. The maximum ARE is 3.03\%, corresponding to the \(\alpha\)-phase volume fraction for the dataset-mean target, while all other AREs are below 1.90\%. Variability across the three seeded runs is also small, with a maximum coefficient of variation of 1.08\%. These results indicate limited sensitivity to starting-set selection for the three seeds examined and close target alignment across in-envelope and extrapolative property regimes.

\subsection{Local Burgers Orientation Relationship (BOR) Analysis}

The conditional diffusion framework is developed for general inverse design of polycrystalline materials, but the present Ti-6Al-4V case has an additional crystallographic consideration. In dual-phase Ti-6Al-4V, neighboring HCP \(\alpha\) and BCC \(\beta\) phases often exhibit the plane-direction alignment associated with the BOR \cite{10.1098/rspa.2014.0881},

\begin{equation}
\{0001\}_{\alpha} \parallel \{110\}_{\beta},
\qquad
\langle 11\bar{2}0\rangle_{\alpha}
\parallel
\langle 111\rangle_{\beta}.
\end{equation}

The initial database was generated with BOR-constrained orientation assignment. However, the parent-child transformation links are not explicitly stored, and the present design pipeline operates on a fixed grain topology at a microstructural snapshot without explicitly modeling its transformation history. The analysis below is therefore a statistical estimate of local crystallographic consistency rather than an exact reconstruction of the original transformation pathway.

For the BOR angle computation, each BOR plane-direction pair is represented as an orthonormal frame. Given a plane normal \(\mathbf{n}\) and an in-plane direction \(\mathbf{d}\), with \(\mathbf{n}\cdot\mathbf{d}=0\), the frame is constructed as
\begin{equation}
\mathbf{F}
=
\left[
\hat{\mathbf{d}},
\widehat{\mathbf{n}\times\mathbf{d}},
\hat{\mathbf{n}}
\right],
\end{equation}
where \(\hat{(\cdot)}\) denotes normalization. The first axis is the in-plane BOR direction, the third axis is the BOR plane normal, and the second axis completes the right-handed orthonormal basis. The Euler angles of each grain are converted into their corresponding Bunge orientation matrices. Let \(\mathbf{R}_i\) be the orientation matrix of \(\alpha\) grain \(i\), and let \(\mathbf{R}_j\) be the orientation matrix of a neighboring \(\beta\) grain \(j\). The corresponding BOR frames are mapped into the sample frame as
\begin{equation}
\mathbf{A}_{ia}
=
\mathbf{R}^{T}_{i}\mathbf{F}^{\alpha}_{a},
\qquad
\mathbf{B}_{jb}
=
\mathbf{R}^{T}_{j}\mathbf{F}^{\beta}_{b},
\end{equation}
where \(a\) and \(b\) index the enumerated BOR frame choices in the \(\alpha\) and \(\beta\) phases, respectively. The relative rotation between an \(\alpha\) BOR frame and a \(\beta\) BOR frame is
\begin{equation}
\Delta\mathbf{R}_{ijab}
=
\mathbf{A}_{ia}^{\top}\mathbf{B}_{jb}.
\end{equation}
The angular deviation from BOR is computed from the trace of the relative rotation matrix as \citep{gallego2015compact,Rowenhorst_2015}
\begin{equation} 
\theta_{ijab} = \cos^{-1} \left[ \frac{ \mathrm{tr} \left( \Delta\mathbf{R}_{ijab} \right) -1 }{2} \right]. \end{equation}
Let \(\mathcal{N}_{\beta}(i)\) denote the set of neighboring \(\beta\) grains of \(\alpha\) grain \(i\). The local BOR angle assigned to \(\alpha\) grain \(i\) is
\begin{equation}
\theta_i^{\min}
=
\min_{j\in\mathcal{N}_{\beta}(i)}
\min_{a,b}
\theta_{ijab}.
\end{equation}
The local BOR consistency of \(\alpha\) grains is quantified using an adopted angular threshold of \(\tau=15^{\circ}\) \cite{Cabibbo2013, Begley2020} as
\begin{equation}
F_{\tau}
=
\frac{1}{N_{\alpha}^{\mathrm{loc}}}
\sum_{i=1}^{N_{\alpha}^{\mathrm{loc}}}
\mathbb{I}
\left[
\theta_i^{\min}<\tau
\right],
\end{equation}
where \(N_{\alpha}^{\mathrm{loc}}\) is the number of \(\alpha\) grains with at least one local neighboring \(\beta\) candidate.

To improve the crystallographic consistency in both target cases, the 32 candidates generated for each input graph are re-ranked using a combined score that augments the property-only matching formulation with an additional local BOR-consistency term:

\begin{equation}
\label{eq:S_BOR}
S_{\mathrm{BOR}}=
w_E\frac{|E-E^*|}{|E^*|}
+w_Y\frac{|Y-Y^*|}{|Y^*|}
+w_{\mathrm{BOR}}\left(1-F_{15}\right),
\end{equation}
where \(F_{15}\) is the local BOR fraction below \(15^\circ\). The weighting factors for property matching remain unchanged with \(w_E=w_Y=1\). The BOR weight is varied over \(w_{\mathrm{BOR}}\in\{0,0.1,0.5,1,10\}\) to investigate the effect of post-generation local BOR-consistency term. 

\begin{figure}[ht]
    \begin{center}
     \centering
     \includegraphics[width=350pt]{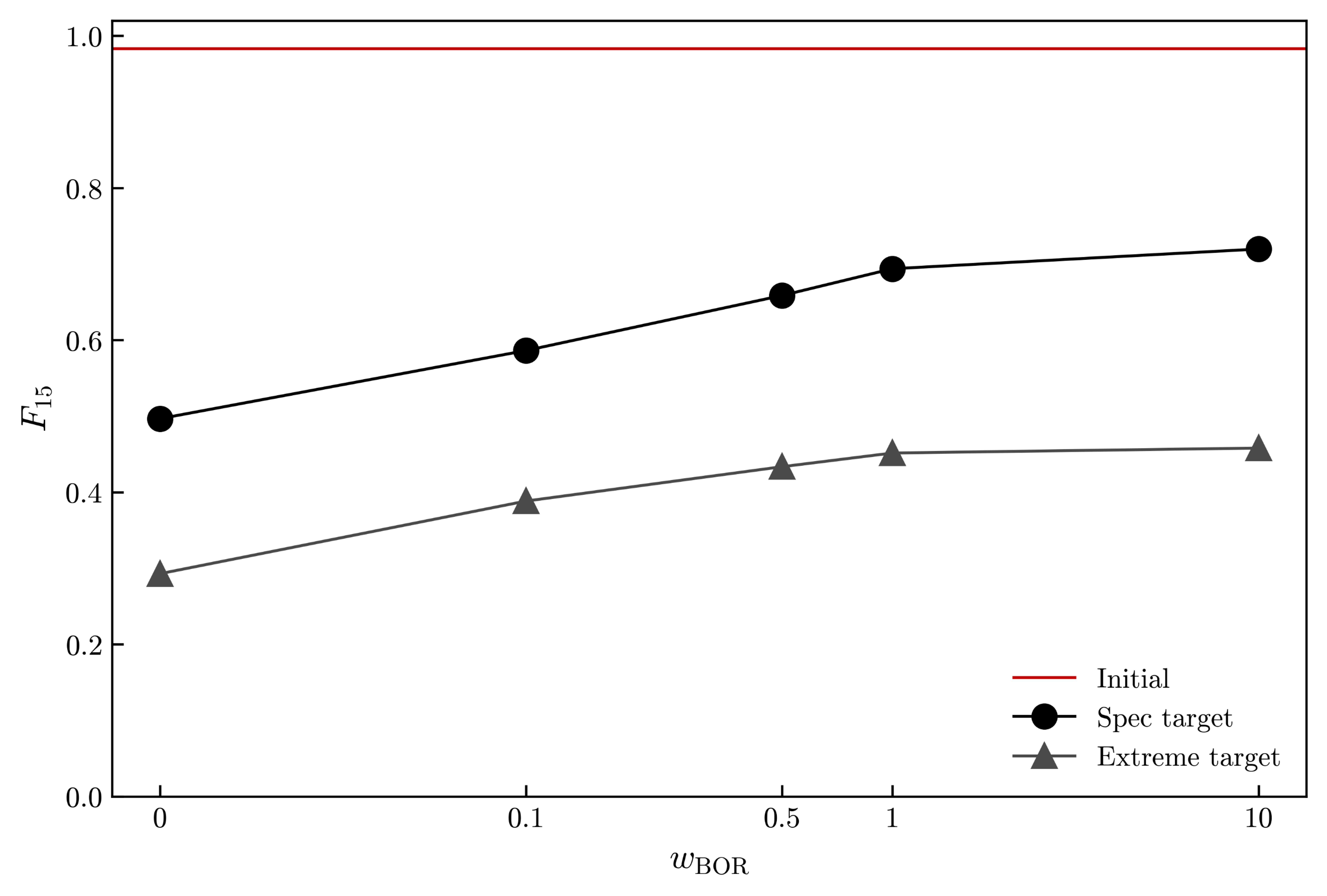}
    \end{center}
    \vspace{-1em}
    \caption{Local BOR fraction of the selected design-to-spec and extreme-property candidates with respect to the BOR weight.}
    \label{fig:BORConsistency}
    \end{figure}

As shown in Fig. \ref{fig:BORConsistency}, the local BOR fraction, \(F_{15}\), is plotted against \(w_{\mathrm{BOR}}\) in logarithmic scale for both the design-to-spec and extreme-property cases. The BOR-constrained input structures have a mean \(F_{15}\) of 0.983. Under property-only ranking, that is \(w_{\mathrm{BOR}}=0\), the selected design-to-spec and extreme-property candidates have mean values of 0.497 and 0.293, respectively. Thus, candidates chosen solely for property agreement are less locally BOR-consistent than the inputs, with a larger reduction for the extreme-property target. This may be associated with its greater departure from the initial data distribution. The database contains no BOR-consistent microstructure near the extreme property target that can provide a nearby reference state for the conditioned reverse process. In addition, the higher target \(\alpha\)-phase fraction reduces local \(\beta\)-phase availability, leaving fewer neighboring \(\beta\) grains with which an \(\alpha\) grain can satisfy the BOR criterion. At \(w_{\mathrm{BOR}}=0.5\), the two means increase to 0.659 and 0.434, corresponding to relative increases of 32.6\% and 48.1\% over property-only ranking. These gains capture 72.6\% and 85.2\%, respectively, of the total improvement observed between \(w_{\mathrm{BOR}}=0\) and 10. At \(w_{\mathrm{BOR}}=1\), the means are 0.694 and 0.452, while increasing the weight to 10 yields 0.720 and 0.458. The marginal improvement therefore diminishes beyond approximately \(w_{\mathrm{BOR}}=1\), particularly for the extreme-property target.

\begin{figure}[ht]
    \begin{center}
     \centering
     \includegraphics[width=250pt]{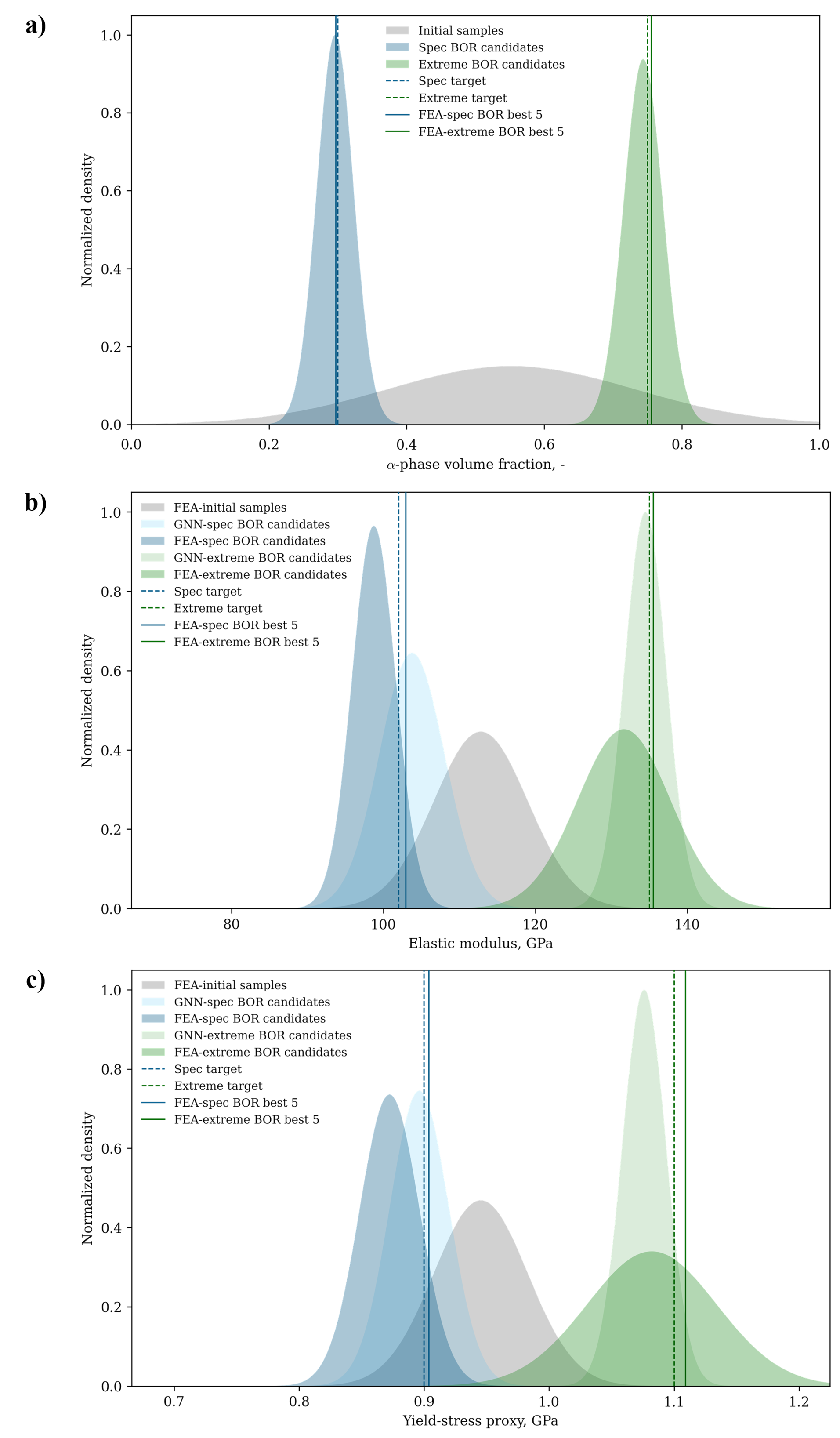}
    \end{center}
    \vspace{-1em}
    \caption{Property distributions of the design-to-spec and extreme-property candidates after BOR-consistency selection: Distributions of a) \(\alpha\)-phase volume fraction, b) elastic modulus, and c) yield-stress proxy for the initial and generated candidate designs.}
    \label{fig:DiffusionDesignBOR}
    \end{figure}

The corresponding property distributions and FEA validation results are shown in Fig.~\ref{fig:DiffusionDesignBOR}. The reported designs are selected using the combined score with \(w_{\mathrm{BOR}}=0.5\) as it captures most of the observed BOR improvement while maintaining substantial emphasis on the property targets. The BOR-consistency-selected distributions continue to overlap the target-property regions, showing that a more BOR-consistent candidate can be identified within each set of 32 candidate designs without a noticeable loss of target agreement. For the best five FEA-validated candidates, the mean design-to-spec properties are \(E=102.94\)~GPa, \(Y=0.904\)~GPa, and \(v_\alpha=0.297\), with AREs of 0.92\%, 0.44\%, and 1.00\%, respectively. The corresponding extreme-property values are \(E=135.53\)~GPa, \(Y=1.11\)~GPa, and \(v_\alpha=0.756\), with AREs of 0.39\%, 0.91\%, and 0.80\%, respectively. The differences in ARE between the BOR-aware and property-only selections are comparable in magnitude, and the property target agreement is therefore considered comparable between the two selection strategies. Accordingly, BOR-aware ranking partially recovers local crystallographic consistency in both cases without a material loss in the target agreement of the best validated candidates. The recovery remains incomplete relative to the input reference, particularly for the extreme-property target. Further improvement may require a larger generated candidate pool or the incorporation of parent-child crystallographic information during the generation process.

\section{Comparison with Random Search and Evolutionary Optimization}

The efficiency of the proposed conditional diffusion framework is compared with two commonly used gradient-free inverse design strategies: random search and evolutionary optimization. The comparison is performed using three initial grain microstructures from the database under the design-to-spec targets \(v_{\alpha}^{*}=0.30\), \(E^{*}=102\)~GPa, and \(Y^{*}=0.90\)~GPa. For random search and evolutionary optimization, the optimization objective includes an explicit \(\alpha\)-phase volume fraction term, because unlike the diffusion model, these methods do not impose the prescribed phase fraction through generation-time phase-fraction projection:

\begin{equation}
\label{eq:OptObjective}
S_{\mathrm{opt}}
=
w_E
\frac{|E-E^*|}{|E^*|}
+
w_Y
\frac{|Y-Y^*|}{|Y^*|}
+
w_{\mathrm{BOR}}
\left(1-F_{15}\right)
+
w_{\alpha}
|v_{\alpha}-v_{\alpha}^{*}|.
\end{equation}
The weighting factors are set to \(w_E=1\), \(w_Y=1\), \(w_{\mathrm{BOR}}=0.5\), and \(w_{\alpha}=0.5\), so that the gradient-free optimization uses the same property matching and BOR-consistency terms as the diffusion candidate selection score, with an additional explicit phase fraction penalty.

\begin{figure}[ht]
    \begin{center}
     \centering
     \includegraphics[width=500pt]{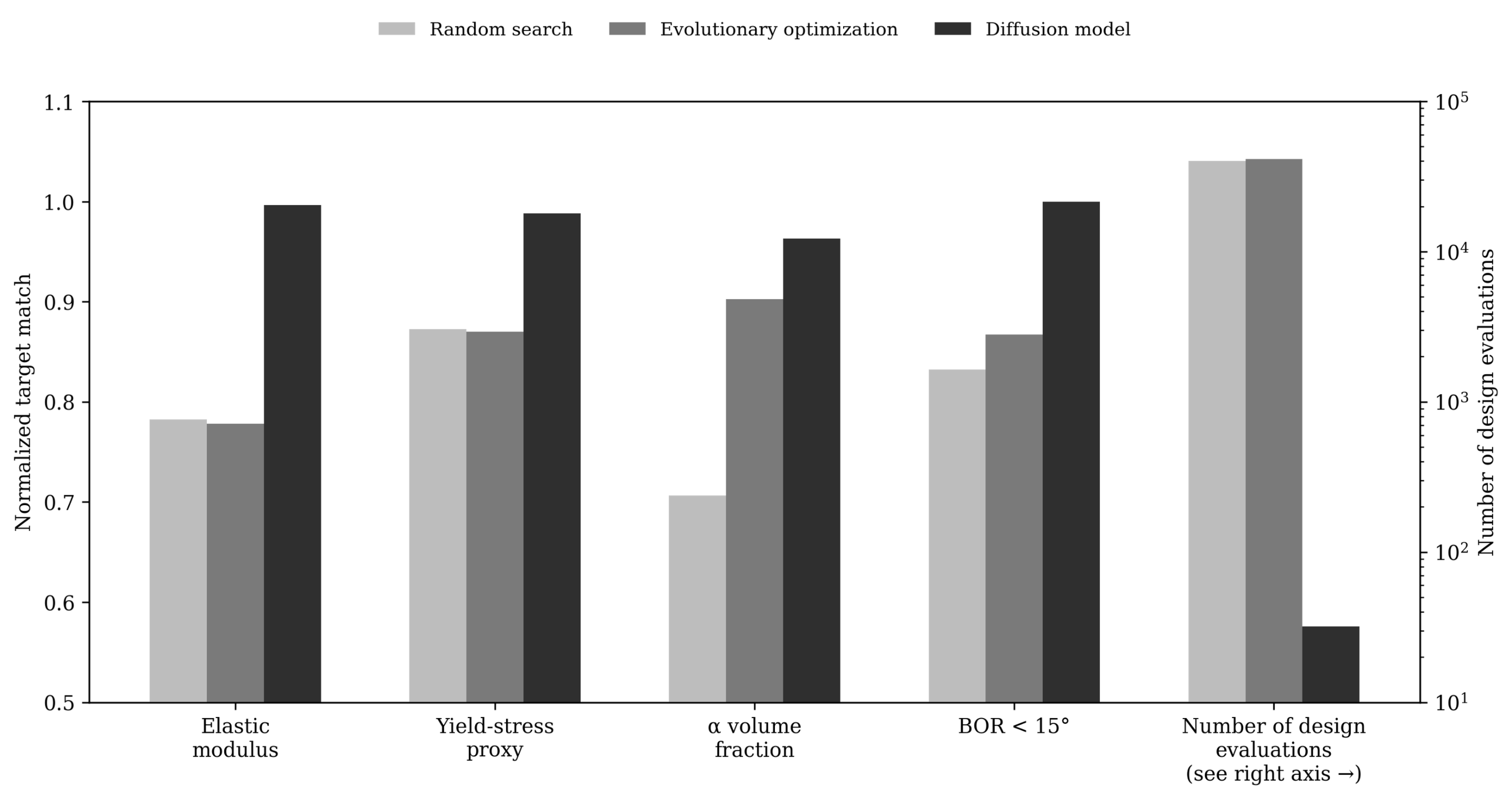}
    \end{center}
    \vspace{-1em}
    \caption{Target alignment, local BOR consistency, and design efficiency of the inverse design methods.}
    \label{fig:methodComparison}
    \end{figure}

Random search and evolutionary optimization are implemented using the same design variables, bounds, and objective function. For each grain, the design variables are the Euler angles and a continuous phase gene in \([0,1]\), which is thresholded at 0.5 to assign the binary phase. Evolutionary optimization is implemented using differential evolution from SciPy \citep{2020SciPy-NMeth} with  a population size multiplier of 3, a maximum of 10 generations, a mutation interval of \((0.5,1.0)\), a recombination constant 0.7, and a convergence tolerance 0.01. All runs reached the prescribed generation limit. The results are therefore interpreted as fixed-budget outcomes rather than converged optimal solution. Random search uniformly samples a prescribed number of independent candidate design vectors within the same design variable bounds. In this work, 40,000 random candidates are evaluated, which is chosen to be comparable to the total number of objective evaluations used by the evolutionary optimization. Fig.~\ref{fig:methodComparison} compares the three inverse design methods in terms of target alignment, local BOR consistency, and design efficiency. For visual comparison, the target alignments for \(\alpha\)-phase volume fraction, elastic modulus, and yield-stress proxy are normalized by their corresponding target values as $1-\frac{|P-P^*|}{P^*}$, where \(P\in\{E,Y,v_\alpha\}\). A normalized target match of 1 indicates an exact target agreement to the design target. The local BOR consistency is normalized by the maximum value among the compared strategies, and the number of design evaluations required to obtain one selected design is plotted on a logarithmic scale. The corresponding numerical values are summarized in Table \ref{tab:methodComparisonTable}.

\begin{table}[htbp]
  \centering
  \tiny
  \caption{Comparison of target alignment, local BOR consistency, and design cost for the inverse design methods.}
    \begin{tabular}{lllllll}
    \toprule
          & \textbf{E (SD), GPa} & \textbf{Y (SD), GPa} & \textbf{\(\alpha\)-vf (SD), -}& \textbf{$\mathbf{F}_{\mathbf{15}}$ (SD), -}& \textbf{Design evaluations (SD), -} & \textbf{Computational time (SD), s} \\
    \midrule
    \textbf{Target} & \multicolumn{1}{r}{\textbf{102.00}} & \multicolumn{1}{r}{\textbf{0.900}} & \multicolumn{1}{r}{\textbf{0.300}} & \textbf{-} &       &  \\
    \midrule
    \textbf{Random search} & 124.21 (0.47) & 1.015 (0.001) & 0.388 (0.012) & 0.577 (0.024) & 40000 (-) & 2800.67 (111.76) \\
    \textbf{Evolutionary optimization} & 124.63 (0.63) & 1.017 (0.009) & 0.329 (0.006) & 0.601 (0.021) & 41404 (5542) & 2363.33 (355.97) \\
    \textbf{Diffusion model} & 102.35 (1.37) & 0.889 (0.008) & 0.311 (0.025) & 0.693 (0.042) & 32 (-) & 24 (0.47)\\
    \bottomrule
    \end{tabular}%
  \label{tab:methodComparisonTable}%
\end{table}%

The conditional diffusion model is substantially more efficient than random search and evolutionary optimization. For each input graph, the conditional diffusion sampling procedure generates 32 candidate designs, and only these 32 generated candidates are subsequently evaluated by the forward GNN for final selection. In contrast, random search and evolutionary optimization require repeated sampling, mutation, and evaluation of trial designs. In the present comparison, the two gradient free optimization strategies require approximately 40,000 design evaluations to obtain one selected design. Thus, the search-based methods require roughly three orders of magnitude more forward property evaluations than the conditional diffusion workflow. 

The computation wall-clock times for the three benchmark problems are also reported in Table \ref{tab:methodComparisonTable}. Training of the diffusion model and all benchmark calculations are performed on the same computation platform equipped with an Intel Core Ultra 9 285K processor operating at 3.70GHz, 64GB of RAM, and an NVIDIA GeForce RTX 5080 GPU with 16GB of memory. The mean runtimes per benchmark problem are 2,801s for random search, 2,363s for evolutionary optimization, and 24s for the diffusion model. Thus, the trained diffusion workflow was approximately two orders of magnitude faster than the search-based methods for the design generation. The reduction in wall-clock time is less pronounced than the reduction in candidate evaluations. Random search and evolutionary optimization require 40,000 and an average of 41,404 candidate evaluations, respectively, corresponding to approximately 1,300 times the 32 candidates evaluated by the diffusion model. In comparison, their runtimes are approximately 117 and 99 times that of diffusion. This difference arises because generating a diffusion candidate is more computationally expensive than evaluating one independently sampled or evolutionarily generated candidate: each diffusion candidate must pass through the complete reverse diffusion before the forward evaluation.

Training the diffusion model introduces a one-time cost of 365s. For a single design problem, conservatively assigning the entire training cost to that problem gives a total diffusion cost of approximately 389s, which remains substantially lower than the corresponding runtimes of 2,801 and 2,363s for random search and evolutionary optimization, respectively. Across the three benchmark problems considered here, the total diffusion cost is 437s, corresponding to an amortized cost of approximately 146s per problem. Because the trained model can subsequently be reused for additional targets and initial graph inputs without retraining, the contribution of this initial training cost decreases as the number of design problems increases, approaching the online sampling cost of approximately 24s per problem. 

In addition to its lower evaluation cost, conditional diffusion also gives the highest alignment for all four reported quantities. Random search and evolutionary optimization give similar elastic modulus and yield-stress proxy alignment, while evolutionary optimization provides noticeably better \(\alpha\)-phase volume fraction alignment and moderately better BOR consistency. 

Overall, the conditional grain-graph diffusion framework provides the closest mean target alignment and highest mean \(F_{15}\) in this benchmark. Unlike random search and evolutionary optimization, which rely on large numbers of trial designs evaluated sequentially or iteratively, the diffusion model learns a conditional generative distribution and directly proposes candidate microstructures near the prescribed target region. Accordingly, the diffusion workflow reduces the number of sample evaluations by approximately three orders of magnitude and the design runtime by approximately two orders of magnitude. This efficiency supports high-throughput grain microstructure inverse design, particularly when the trained model is reused across multiple design queries.

\section{Conclusions}

In this work, a conditional grain-graph diffusion framework is developed for the inverse design of dual-phase Ti-6Al-4V microstructures. The proposed framework represents each microstructure as a grain graph, where grains are encoded as nodes and grain boundaries are encoded as edges. Based on this graph representation, a GNN is first constructed as a surrogate model for forward stress prediction, and a conditional graph DDPM is then developed for inverse design by denoising node-level phase and crystallographic orientation states under prescribed property targets.

The grain GNN shows accurate forward stress prediction performance, particularly in the low strain regime. The model achieves an \(R^2\) of 0.95 with an MAE of 0.003~GPa at 0.25\% strain, and maintains good accuracy up to 0.75\% strain with an \(R^2\) of 0.93 and an MAE of 0.008~GPa. Due to the increased nonlinearity of the plastic response, the prediction accuracy decreases after the onset of plasticity, reaching an \(R^2\) of 0.78 and an MAE of 0.012~GPa at 1.50\% strain. Compared with the voxel-based CNN benchmark, the grain GNN achieves comparable prediction accuracy using a more compact and efficient graph-based microstructure representation. For the representative microstructure with 38 grains, a \(32^3\) voxelized CNN input contains 32,768 voxel locations before accounting for feature channels. In contrast, the graph representation encodes the same microstructure using only 38 grain nodes and 131 undirected grain boundary edges. Compared with the baseline grain GNN, the proposed model shows improved accuracy thanks to the addition of grain boundary property features, node and edge feature embeddings, and multi-statistic graph pooling.

The conditional diffusion model is evaluated across four target regimes using three independently seeded starting sets of 50 microstructures for each regime. The generated candidates approach the prescribed targets within and beyond the property envelope of the initial microstructures. The diffusion model also shows limited sensitivity to starting-set selection over the three seeds examined. For the two primary design cases, selected microstructure designs are further validated by FEA, and the top five FEA-validated candidates show close agreement with the prescribed targets. As the \(\alpha\) and \(\beta\) phases in Ti-6Al-4V are commonly related through the BOR, local BOR consistency is evaluated as a crystallographic consistency metric for the generated designs. The candidates selected using property-only ranking exhibit lower local BOR consistency than the BOR-constrained input microstructures, with a greater reduction in the extreme-property case. This reduction can be attributed in part to the absence of a nearby BOR-consistent input microstructure to guide reverse diffusion toward the out-of-envelope target. In addition, the higher target \(\alpha\)-phase volume fraction leaves fewer neighboring \(\beta\) grains available for local BOR matching. Introducing a BOR-consistency term into the candidate selection score partially recovers local crystallographic consistency in both cases, increasing the mean BOR consistency relative to property-only ranking by up to 44.9\% and 56.4\% for the in- and out-of-envelope targets, respectively, while retaining close property target agreement with an ARE of 1.0\%. 

The efficiency of the diffusion framework is compared with random search and evolutionary optimization. For the same design task, the diffusion model shows the best overall target alignment and design efficiency. The diffusion workflow requires only 32 generated candidates to be evaluated for final selection, whereas random search and evolutionary optimization require approximately 40,000 objective evaluations. This corresponds to approximately three orders of magnitude fewer terminal evaluations and, for the tested implementations, approximately two orders of magnitude lower runtime. These gains arise because the diffusion model learns a conditional generative distribution and directly proposes candidate microstructures near the target region, rather than relying on large numbers of iterative trial evaluations.

Overall, this work demonstrates that conditional grain-graph diffusion provides an effective computational framework for the property-guided inverse design of polycrystalline microstructures. By integrating a compact grain-graph representation, a GNN forward surrogate, and a conditional graph DDPM, the proposed framework enables multi-target microstructure design with substantially fewer candidate evaluations than conventional search-based approaches. For Ti-6Al-4V specifically, the results further demonstrate that crystallographic constraints, such as local BOR consistency, can be incorporated through post-generation candidate selection to balance property-target alignment and crystallographic consistency. However, more complete recovery of BOR consistency may require a larger candidate pool or the explicit incorporation of parent-child crystallographic information into the generative process.

    \printcredits

    \section*{Acknowledgment}
    This work was supported by the Natural Sciences and Engineering Research Council of Canada (NSERC) Discovery Grant (RGPIN-2026-06958), New Frontiers in Research Fund-Exploration (NFRFE-2019-00753), Digital Research Alliance of Canada. Y. Zou acknowledges the Canada Research Chair in Materials and Manufacturing for Extreme Environments.
    
	\section*{Declaration of interests}
	The authors declare that they have no known competing financial interests or personal relationships that could have appeared to influence the work reported in this paper.

    \section*{Declaration of generative AI and AI-assisted technologies in the manuscript preparation process}
    During the preparation of this work the authors used Claude in order to improve readability and language. After using this tool/service, the authors reviewed and edited the content as needed and take full responsibility for the content of the published article.

	\section*{Data availability}
    The code related to this work is publicly available at \url{https://github.com/YuhengZhou-McGill/Conditional_Grain_Graph_Diffusion.git}. The repository contains a representative subset of 200 Ti-6Al-4V grain graphs with corresponding labels. The complete microstructure database is available upon reasonable request.

	\bibliographystyle{elsarticle-num-names}
	\bibliography{cas-refs}

\end{document}